\documentclass[amsmath,amssymb,aps,floats,prl,twocolumn,showpacs,unsortedaddress,superscriptaddress]{revtex4}
\usepackage{graphicx}
\usepackage{amssymb}
\usepackage{amsmath}
\usepackage{epstopdf}
\usepackage{color}
\usepackage[normalem]{ulem}
\def \b {\beta}
\def \d {\delta}

\def \ve {\varepsilon}

\def \g {\gamma}
\def \G{\Gamma}
\def \k {\kappa}

\def \o {\omega}
\def \O {\Omega}
\def \t {\theta}

\def \s {\sigma}

\def \dag {\dagger}

\def \p {\partial}

\def \del {\nabla}

\def \curl {\nabla\times}
\def \delk {\nabla_{\mbf k}}

\def \apx {\approx}

\def \dag {\dagger}

\def \iint {\int_{-\infty}^{\infty}}
\def \pint {\int_{0}^{\infty}}

\def \hcint {\int_{0}^{\pi}}

\def \intvk {\int_{\mbf k}}

\def \uar {\uparrow}
\def \dar {\downarrow}

\def \rar {\rightarrow}

\def \ra {\rangle}
\def \fr {\frac}
\def \lf {\left}
\def \ri {\right}

\newcommand{\ket}[1]{|#1\ra}

\def \bece {\begin{center}}
\def \ence {\end{center}}
\def \beeq {\begin{equation}}
\def \eneq {\end{equation}}
\def \beal {\begin{aligned}}
\def \enal {\end{aligned}}
\def \bega {\begin{gathered}}
\def \enga {\end{gathered}}
\def \benu {\begin{enumerate}}
\def \ennu {\end{enumerate}}
\def \beit {\begin{itemize}}
\def \enit {\end{itemize}}
\def \bede {\begin{description}}
\def \ende {\end{description}}
\def \betb {\begin{tabular}}
\def \entb {\end{tabular}}
\def \bear {\begin{array}}
\def \enar {\end{array}}
\def \tbf {\textbf}
\def \tit {\textit}
\def \mbf {\mathbf}

\def \mrm {\mathrm}

\def \bsb{\boldsymbol}

\newcommand{\bea}{\begin{eqnarray}}
\newcommand{\ena}{\end{eqnarray}}
\newcommand{\bee}{\begin{equation}}
\newcommand{\ene}{\end{equation}}

\newcommand{\bfig}{\begin{figure}}
\newcommand{\efig}{\end{figure}}

\newcommand{\black}{\color{black}}

\newcommand{\mini} {\mathrm{min}}

\newcommand{\sxx} {\sigma_{xx}}
\newcommand{\sxy} {\sigma_{xy}}
\newcommand{\rxx} {\rho_{xx}}
\newcommand{\rxy} {\rho_{xy}}
\newcommand{\ep} {\epsilon}
\newcommand{\delr} {\Delta\rho}

\newcommand{\hallang} {\tan\theta_H} 

\newcommand{\nbsb} {NbSb$_2$}
\newcommand{\tasb} {TaSb$_2$}


\begin{document}

\title{ Non-saturating large magnetoresistance in semimetals }

\author{Ian A. Leahy }
\author{Yu-Ping Lin}
\affiliation{Department of Physics, University of Colorado, Boulder, CO 80309, USA}%
\author{Peter E. Siegfried }
\affiliation{Department of Physics, University of Colorado, Boulder, CO 80309, USA}%
\author{Andrew C. Treglia}
\affiliation{Department of Physics, University of Colorado, Boulder, CO 80309, USA}%
\author{Justin C. W.  Song}
\affiliation{Division of Physics and Applied Physics, Nanyang Technological University, Singapore 637371}
\author{Rahul M. Nandkishore}
\affiliation{Department of Physics, University of Colorado, Boulder, CO 80309, USA}
\affiliation{Center for Theory of Quantum Matter, University of Colorado, Boulder, CO 80309, USA}
\author{Minhyea Lee}
\email{minhyea.lee@colorado.edu}
\affiliation{Department of Physics, University of Colorado, Boulder, CO 80309, USA}%
\affiliation{Center for Experiments on Quantum Materials, University of Colorado, Boulder, CO 80309, USA}

\date{\today}

\begin{abstract}
The rapidly expanding class of quantum materials known as {\emph{topological semimetals}} (TSM) display unique transport properties, including a striking dependence of resistivity on applied magnetic field, that are of great interest for both scientific and technological reasons. 
So far  many possible sources of  extraordinarily large non-saturating magnetoresistance   have been proposed. %-- such as  inhomogeneity or disordered-induced  current misalignment  in the classical models and   the pure quantum  effects  including Landau levels and linear dispersion originated from the topological nature of  carriers, to name  a few. Nonetheless,  
However, experimental signatures that  can identify or discern the dominant mechanism and connect to available theories are scarce.  
Here we present the  magnetic susceptibility ($\chi$),  the tangent of the Hall angle ($\hallang$) along with magnetoresistance in four different non-magnetic semimetals with high mobilities, NbP, TaP, \nbsb~ and \tasb, all of which exhibit  non-saturating large MR.   We find that  the  distinctly different temperature dependences, $\chi(T)$  and  the values of $\hallang$  in phosphides and antimonates  serve as  empirical criteria to  sort the MR from different origins: NbP and TaP being uncompensated semimetals with linear dispersion, in which the non-saturating magnetoresistance arises due to guiding center motion, while  \nbsb~and \tasb~ being  {\it compensated} semimetals, with a magnetoresistance emerging  from nearly perfect charge compensation of two quadratic bands.
 Our results illustrate how a combination of magnetotransport and susceptibility measurements may be used to categorize the increasingly ubiquitous non-saturating large magnetoresistance  in TSMs. 
\end{abstract}

\maketitle

%{\emph {Significance Statement}} 
%The intensive recent investigations of topological semimetals have revealed a large number  of semimetal compounds that exhibit very large non-saturating magnetoresistance.  Multiple  mechanisms for this magnetoresistance phenomenon have been theoretically proposed, but experimentally it is unclear how to identify which mechanism is responsible in a particular sample, or how to make a clean connection between experimental observations and theoretical models.   
%Our results  show  that the magnetic susceptibility and the tangent of the Hall angle  successfully capture the  fundamental differences  in seemingly similar  non-saturating large  magnetoresistance, where  charge compensation, energy dispersion and the roles of disorder  are markedly distinct, and provide empirical templates to characterize the origins of the extraordinary magnetotransport  properties in the newly discovered topological semimetals  and beyond.

{\emph {Introduction }} Magnetoresistance (MR) and the Hall effect are versatile experimental probes in exploring electronic properties of materials, such as carrier density, mobility and the nature of scattering and disorder. In typical non-magnetic and semiconducting materials, the MR increases quadratically with applied transverse magnetic field  and saturates to %at 
a constant value when the product of the applied field and the mobility ($\nu$)  approaches unity. 
Non-saturating MR is commonly attributed to the semiclassical two-band model, where electron-like and hole-like carriers are nearly compensated \cite{AshcroftMerminbook}, resulting in  rich magnetotransport characteristics that are  strongly  temperature ($T$) and applied transverse magnetic field ($H$) dependent  in non-magnetic compounds. 
A flurry of interest in non-saturating, $H$-linear MR \cite{Xu1997, Yang1999, Lee2002, Husmann2002} in narrow gap semiconductors led to two main theoretical accounts: (i) a two-dimensional simple 4-terminal resistor network model, where strong disorder or inhomogeneity of the sample manifest as charge and mobility fluctuations \cite{Parish2005,Kisslinger2017} and (ii) the so-called  `quantum linear MR' which emerges in systems with linear band crossings  when the lowest Landau level is occupied \cite{Abrikosov1998}. The former approach has provided a basis to engineer large magnetotransport responses via macroscopic inhomogeneities or disorder \cite{Solin2002,Branford2005, Rosenbaum2008}. Meanwhile, the latter has remained rather elusive until recently. 

Interest in non-saturating very large MR  has exploded following the discovery of topological semimetals. These materials are regularly reported  to exhibit record high non-saturating MR, known as extreme magnetoresistance (XMR) with unusually high mobilities for bulk systems \cite {Narayanan2015, Liang2015, Luo2016TaAs,Huang2015,LvPRX2015,Shekhar2015,JunfengHe2016} and relatively low residual resistivity $\rho_0$. %:
The proximity of the chemical potential to  the  charge neutrality point in semimetals  allows  the  generic quadratic two band model to describe the MR and the Hall effect in reasonable levels  \cite{Ali2014, Luo2015WTe, Zeng2016,Xu2017YSb, Yuan2016}.  
However, a two band model of this form generically predicts a magnetoresistance that is {\it quadratic} in applied fields, whereas the materials frequently exhibit a magnetoresistance linear in applied field. While various theoretical proposals for $H$ linear magnetoresistance have been advanced (see e.g. \cite{Polyakov1986,  Abrikosov1998, Song2015, Mirlin2015}) the origins of extreme magnetoresistance in topological semimetals remain unclear.  
As non-saturating, large MR becomes more ubiquitous, it becomes particularly urgent %pertinent 
to identify a set of distinct attributes that enable  the delineation of \black%to delineate
 their origins. 

In this article, we systematically %comparatively 
examine the low field diamagnetic susceptibility ($\chi$), the transverse MR, and the Hall effect as a function of $T$ and $H$ in 4 different semimetals with high mobility ($\nu \ge 10^4$ cm$^2$/Vs) and very large non-saturating MR -- NbP, TaP (phosphides), \nbsb, and \tasb (antimonates). Characteristic parameters related to magnetic transport are summarized in Table \ref{table1}. 
\begin{table}[b]
\begin{center}
\begin{tabular}{ccccc} 
\hline\hline
 &$\tan^2\theta_H$  &$\nu$ (T$^{-1}$) &$\rho_0$ ($\mu\Omega$  cm)& $\delr/\rho_0$\\ %&RRR   \\
\hline\hline
NbP& 7.6 &  99 &  0.5&561 \\ %& 30\\
TaP&  5.8& $3.5\times10^3$ &0.2& 20200 \\ %& 90\\
\hline
\nbsb&  $\le 10^{-4}$ & 1.9-2.5 &0.1 &27800  \\%& 600\\
\tasb& $\le 10^{-4}$   &  2.2-4.3 & 0.1 &5560 \\%& 290\\
\hline\hline
\end{tabular}
\caption {\small  Summary of magnetotransport  data. Residual resistivity $\rho_0$ at zero field is reported at $T= 2 $ K and  $\delr/\rho_0=(\rho(H)-\rho_0)/\rho_0$ and $\tan^2\theta_H$  at $0.3$ K and   $\mu_0H = 15$ T.  }
\label{table1}
\end{center}
\end{table}

We present two different types of non-saturating large MR identified by the temperature ($T$) dependence of diamagnetic susceptibility, $\chi(T)$ and the $H$ dependence of the Hall angle, 
$\tan\theta_H=\frac{\rxy}{\rxx}=\frac{\sxy}{\sxx}$,
where $\rxx$ and $\rxy$  are longitudinal and Hall resistivity respectively and $\sxx$ and $\sxy$ are corresponding conductivities.

One type of MR originates from the presence of smooth disorder that governs guiding center motion of charge carriers. The linear $H$-dependence of this type arises from the squeezed trajectories of carriers in semi-classically large magnetic fields $\nu B \ge 1$ (easily achieved in linearly dispersing topological semimetals, see e.g., Table~\ref{table1}), and does not require the involvement of multiple bands for the charge compensation.
The other type of MR comes from charge compensation in the two band model and it accompanies  other  transport and magnetic characteristics within the conventional frame work.

%%% FIG 0
\begin{figure}[t!]%[ht]
\begin{center}
\includegraphics[width=0.9\columnwidth]{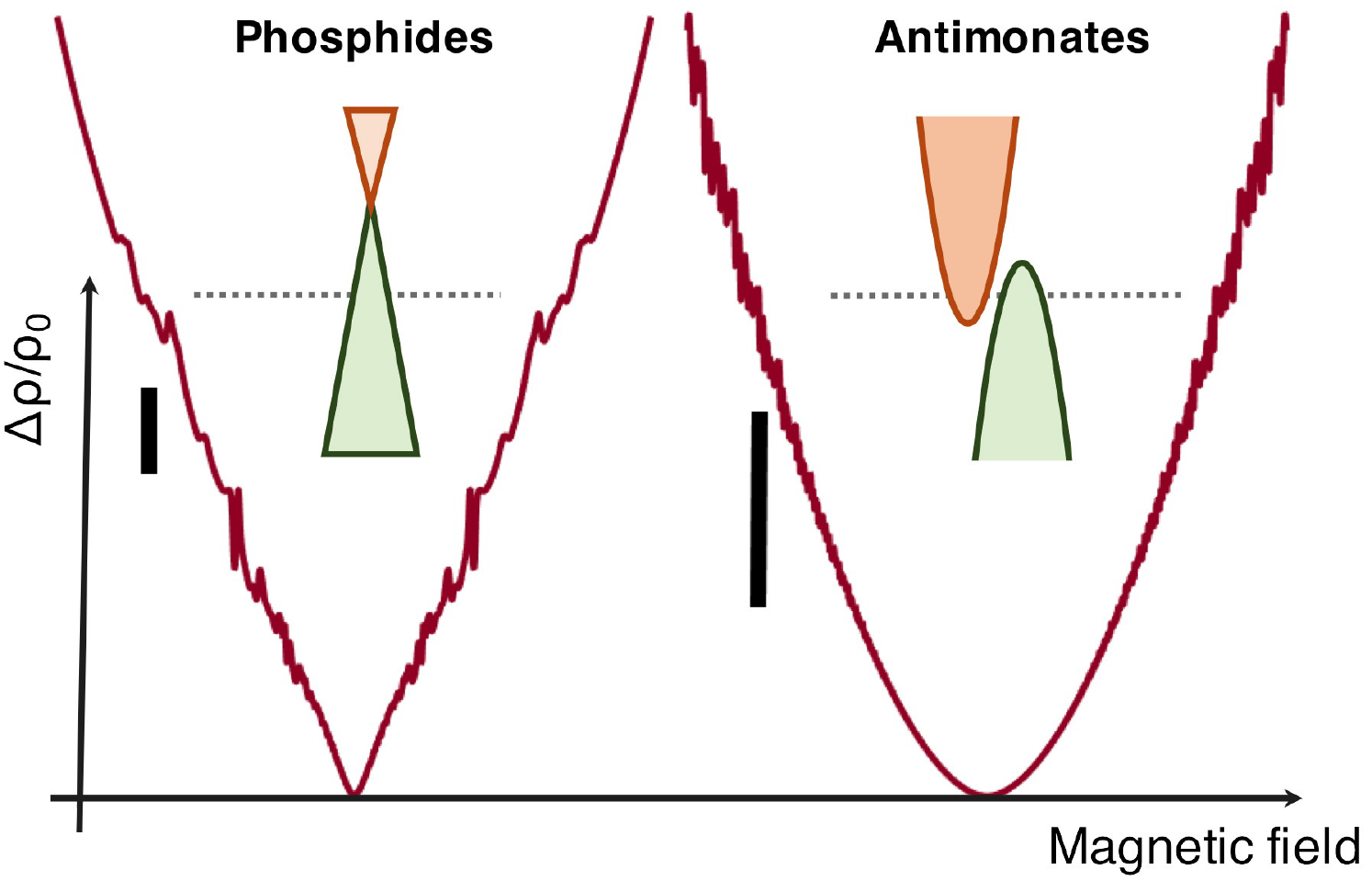}
\caption{\small Schematically depicted non-saturating  MR phenomena and representative energy dispersions for phosphides [TaP] (left) and antimonates [\tasb] (right). 
Phosphides' MR  is characterized by quasi-linear to linear transition as $H$ increases, while antimonates' by persistent quadratic $H$ dependence, 
arising from semiclassical charge compensation. 
%(See text). 
Each bar indicates $\Delta\rho/\rho_0 = 5\times 10^5$ \% up to $\mu_0H = \pm 31$ T at $T=0.3$ K. 
}
\label{schematic}
\end{center}
\end{figure}

Using a combination of magnetic susceptibility and magnetotransport measurements to interrogate the different facets of magneto-response, we are able to categorize the phosphides into the former and the antimonates the latter.  Our results can be summarized  as follows  and depicted in Fig. \ref{schematic}. 
(1) In the phosphides, the magnitude of $\hallang$ saturates to a $H$-independent constant at low temperatures when $H>H_S \simeq 8$ T, while $\rxx(T)$ has a peculiar $H$ dependent non-monotonic form.  
The measured MR defined as $\delr/\rho_0\propto H^\alpha$ at low $T$ exhibits a crossover from quasi-linear [$\alpha \sim 1.5 \pm 0.1$]  to linear [ $\alpha \sim 1.0\pm 0.1$], where the crossover field, $H_S$ is set  by the scale at which $\hallang (H)$  saturates.  
In $H>H_S$, MR remains linear in $H$ up to $\mu_0H = 31$ T, the highest applied field in this study.  Finally, $\chi(T)$'s for the phosphides exhibit a pronounced minimum at $T_{\mini}$. 
  All of these features can be explained  if we assume that the phosphides are semimetals with linear dispersion, even without invoking compensation, and that the magnetoresistance arises due to guiding center motion [see e.g. \cite{Song2015} for a recent discussion]. Moreover, $\chi(T)$ allows to extract doping levels relative to the charge neutrality point as fit parameters.  
(2) Meanwhile, in the antimonates, the Hall angle remains close to zero ($<10^{-2}$) at all accessible fields in this study. The magnetoresistance is nearly quadratic in $H$ from room temperature down to $T=0.3$ K, obeying Kohler's rule. The field dependence of the Hall resistivity strongly deviates  from linearity in the antimonates. The diamagnetic susceptibility for the antimonates is mostly $T$-independent.  
 These features of 
 the antimonates are archetypical for compensated semimetals  with usual quadratic bands.

%\blue
Our finding is well-consistent with  existing electronic structure calculations : NbP and TaP  have  been studied thoroughly  via first principle calculations and photoemission studies  \cite{ChiChengLee2015, SuYangXu2015, Shekhar2015}, where multiple Weyl nodes were identified in vicinity of Fermi energy. The calculations for  \nbsb~ and \tasb~are also consistent with our picture of  nearly compensated semimetals \cite{ChenchaoXu2016}, yet the photoemission studies are not yet avaialble  for the antimonates. 
%ARPES data is not yet available for the antimonates. 
\black

{\emph {Methods }}  Single crystals of NbP, TaP, \nbsb, and  \tasb~were grown using the chemical vapor 
transport method following known synthesis procedure \cite{Wang2014, YukeLi2016, Shekhar2015,Zhang2017}. Standard electrical contacts were made directly on  single crystals using Ag paint (Dupont 4966) with contact resistance ranges in $\le 1-2~\Omega$. The magnetotransport measurements were performed with applied field perpendicular to the direction of current on the plane up to $31$T down to  $0.3$ K. Magnetic susceptibilities  of the samples were measured by the Magnetic Properties Measurements System by Quantum Design.

%%% FIG 1
\begin{figure}[t!]%[ht]
\begin{center}
\includegraphics[width=\columnwidth]{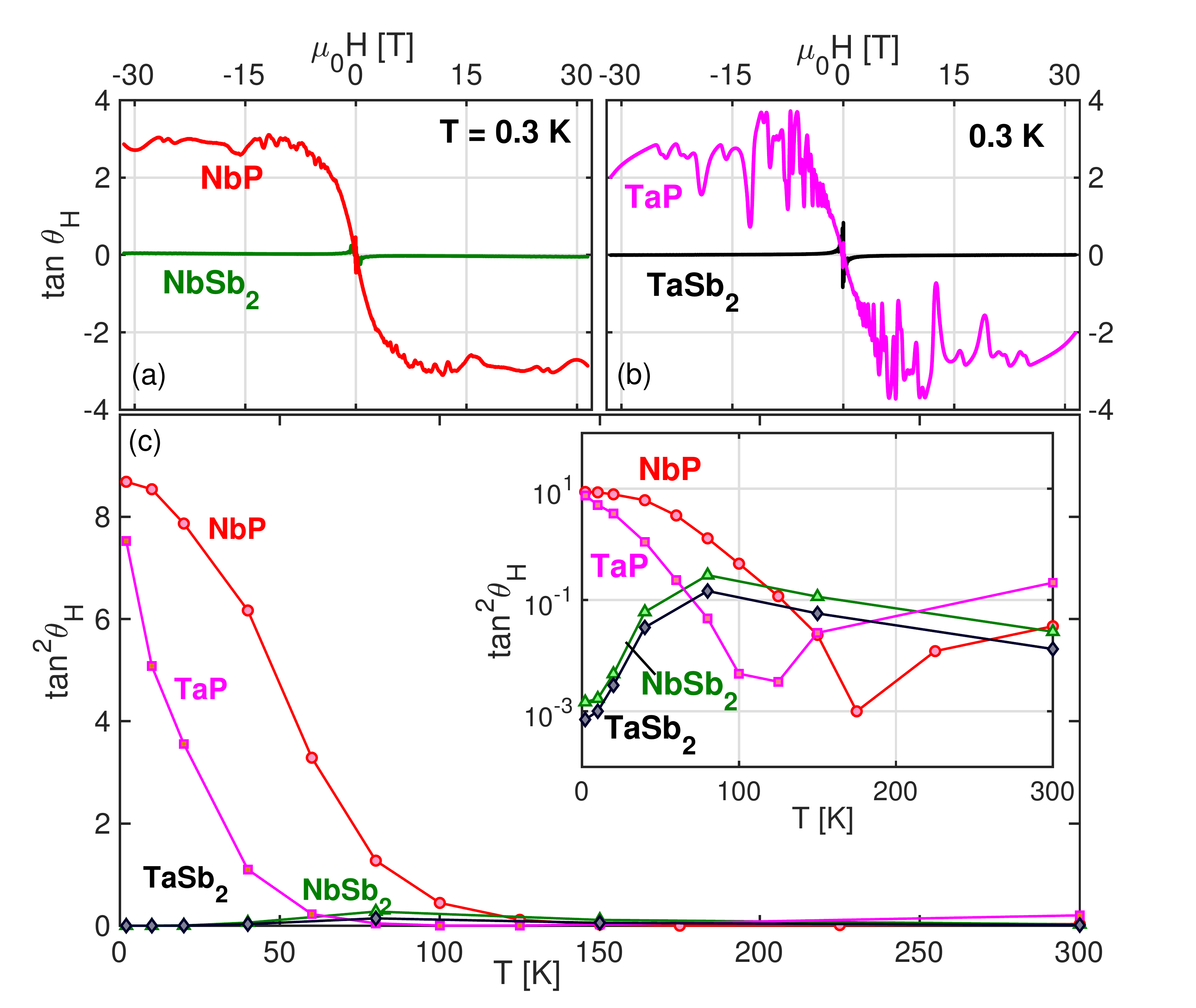}
\caption{\small (a) The Hall angle $\hallang$ for NbP (red) and \nbsb~(green)  as a function of $H$ and 
(b) for TaP (magenta) and \tasb~(black), measured at  $T= 0.3$ K  
Strong quantum oscillations in phosphides results in spike-like features. 
(c) $\tan^2 \theta_H$ measured at $\mu_0H =7$ T as a function of $T$ (Inset) $\tan^2\theta_H$  vs $T$ plotted in $\log$-$\log$ scale, \nbsb~and \tasb~ data are clearly resolved. 
}
\label{hallang}
\end{center}
\end{figure}

{\emph {Results}} 

{\it Magneto-transport} Fig. \ref{hallang} (a) and (b) display $\hallang$  as a function of $H$ at $T=0.3$ K. 
In the high field limit, the phosphides and the antimonates show sharply contrasting behavior: $\hallang$ for NbP and TaP reaches large values saturating to around 2.5 when $H>H_S\simeq 8$ T, while it remains two orders of magnitude smaller for the antimonates (except near zero field). Strong Schubnikov-de-Haas  oscillations are apparent in both $\rxx$ and $\rxy$, that  generate spike-like features in the phosphides. 
In the antimonates, quantum oscillations emerge as well, but only in higher fields and the magnitudes  are \black much smaller due to the smaller Fermi surfaces of the antimonates \cite{Arnold2016, Klotz2016,ChenchaoXu2016}.

The phosphides and antimonates also display contrasting $T$-dependence in Hall angle. Fig.~\ref{hallang}(c) shows the temperature dependence of $\tan^2\theta_H$ measured at $\mu_0H=7$ T. Strikingly, $\tan^2\theta_H$ rises rapidly above unity with decreasing $T$ around 100 K and $60$ K, for NbP and TaP respectively. In contrast, the antimonates behave in the opposite fashion: upon decreasing temperature, $\hallang$ rapidly decreases, giving values two orders of magnitude smaller than the phosphides [Inset of Fig.\ref{hallang}(c)]. 
We note that small Hall angles are frequently found in conventional metals and semimetals \cite{Hurdbook}, as well as in a wide range of XMR materials with high mobilities for both  holes and electrons \cite{Ali2014,Xu2017YSb}. 
 
The field dependence of $\hallang$ plays a deciding role in determining magnetoresistance. For example, %It is important to note that the 
a field-independent $\hallang$ indicates the field dependence of $\rxx$ and $\rxy$ should have the same functional form. In uncompensated systems, $\rxy$ has an $H$-linear Hall contribution (i.e. $\rxy=\mu_0R_HH$ where $R_H$ is the normal Hall coefficient), allowing a field independent $\hallang$ and therefore a non-saturating $H$-linear $\rxx$. This is exactly what %In fact 
we observe %that 
in the phosphides which exhibit $H$-linear $\rxy$ as well as $H$-linear $\rxx$, see below.

Furthermore, large $\hallang$  can act to suppress resistivity and morph its $T$-dependence.
To demonstrate this, %In order to make this more apparent,
we express $\rxx$ in terms of $\hallang$ and $\sxx$, 
\bee 
\rxx =  \frac{\sxx}{\sxx^2 + \sxy^2}=\rxx'\Big(\frac{1}{1+\tan^2\theta_H}\Big),
\label{rxxinv}
\ene
where $\rxx'\equiv\frac{1}{\sxx}$. As evident in Eq.~\ref{rxxinv}, when $\hallang\ge 1$ the inverse relation between $\sxx$ and $\rxx$ no longer holds.
%in the limit of $\hallang\ge 1$.

%%% FIG2
\begin{figure}[t!]
\begin{center}
\includegraphics[width=0.95 \linewidth]{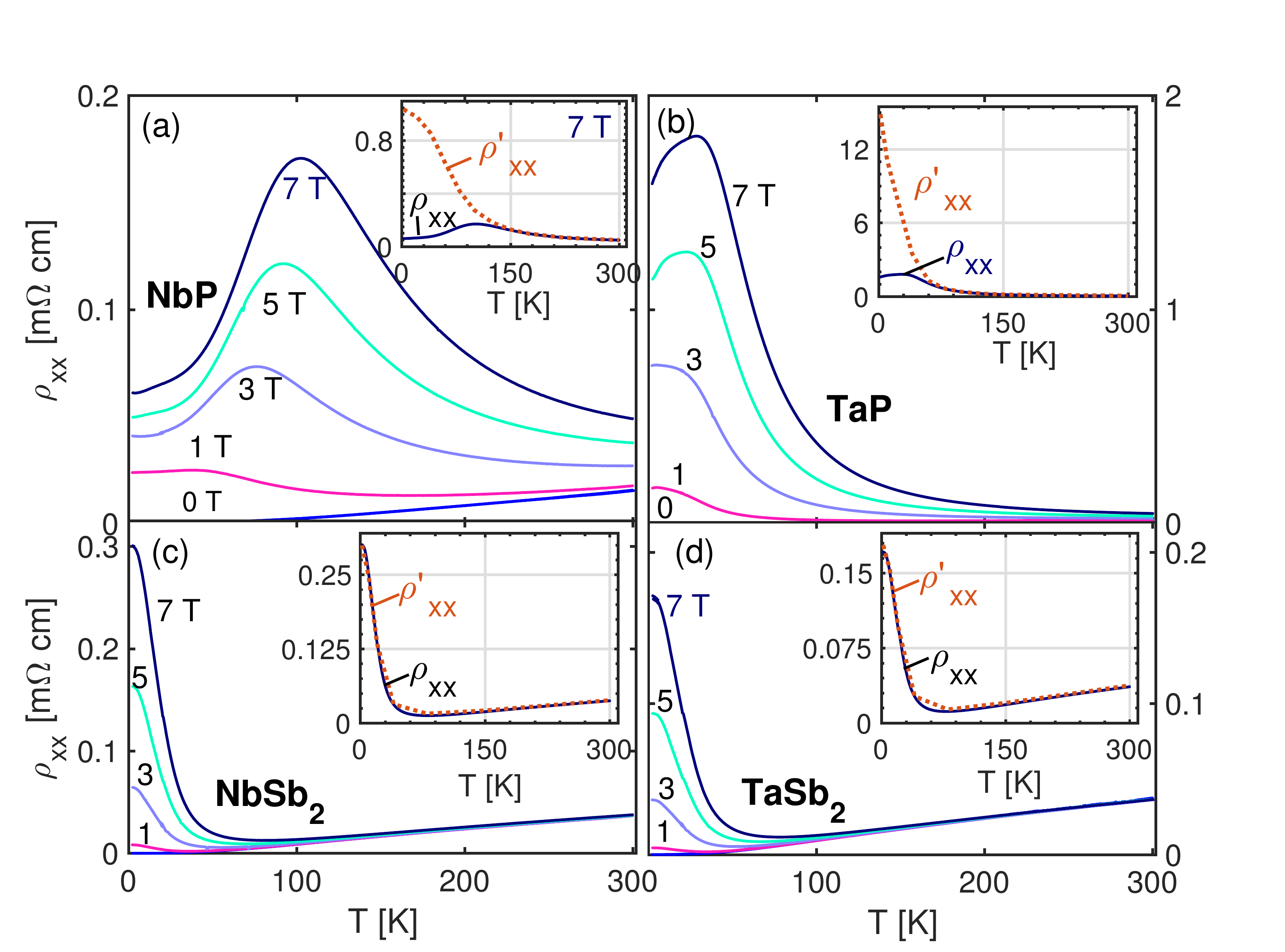}
\caption{\small $\rxx(T)$ at different  $H$'s are shown in (a) NbP, (b) TaP, (c) \nbsb~ and (d) \tasb.  
$\rxx'(T)$ at $\mu_0H = 7$ T, defined in Eq.(\ref{rxxinv}), is plotted in each inset. }
\label{MRT}
\end{center}
\end{figure}

%\blue
The effect of large
$\hallang$ magnitude on $\rxx(T)$ is particularly pronounced for NbP and TaP in Fig.~\ref{MRT}(a-b) where  the $T$ dependence of $\rho_{xx}$ is plotted. 
Monotonic metallic T dependence switches  to non-monotonic behavior  as field increases, with a peak at a temperature that coincides with the onset of rapid increase of $\hallang$ [Fig. 1(c)]. 
%These display a non-monotonic behavior 
%possessing a peak at a temperature that coincides with the onset 
%of rapid increase of $\hallang$ [Fig. 1(c)]. 
\black
Crucially, $\rxx'$ [as defined in Eq. (\ref{rxxinv})] plotted as broken lines in the inset, deviates significantly from the measured $\rxx$, reflecting the large values of $\hallang$ and the dominant role $\hallang$ has in $\rho_{xx}(T)$. 

\nbsb~ and \tasb, however, display contrasting behavior plotted in lower two panels Fig. \ref{MRT}(c,d). We first note that %we plot the corresponding data for \nbsb~ and \tasb,  and 
$\rxx(T)$ initially exhibits a slight decrease in $\rxx$ as $T$ is lowered until the sudden rise, mimicking  a metal-insulator-like transition. This behavior is commonly observed in many XMR materials and $\rxx$ continues increasing as $T$ is lowered further. In contrast to  phosphides  of Fig. \ref{MRT}(a,b), \nbsb~ and \tasb, however, display $\rxx\approx 1/\sxx$ as reflected by the near overlay of the broken-lines and solid lines in the inset. This is consistent with a small $\hallang \ll 1$. For all samples, the rapid rise in $\rxx'$ at low temperatures corresponds to a plummeting $\sxx$.

%%% FIG 3
\begin{figure}[t!]
\begin{center}
\includegraphics[width=0.95 \linewidth]{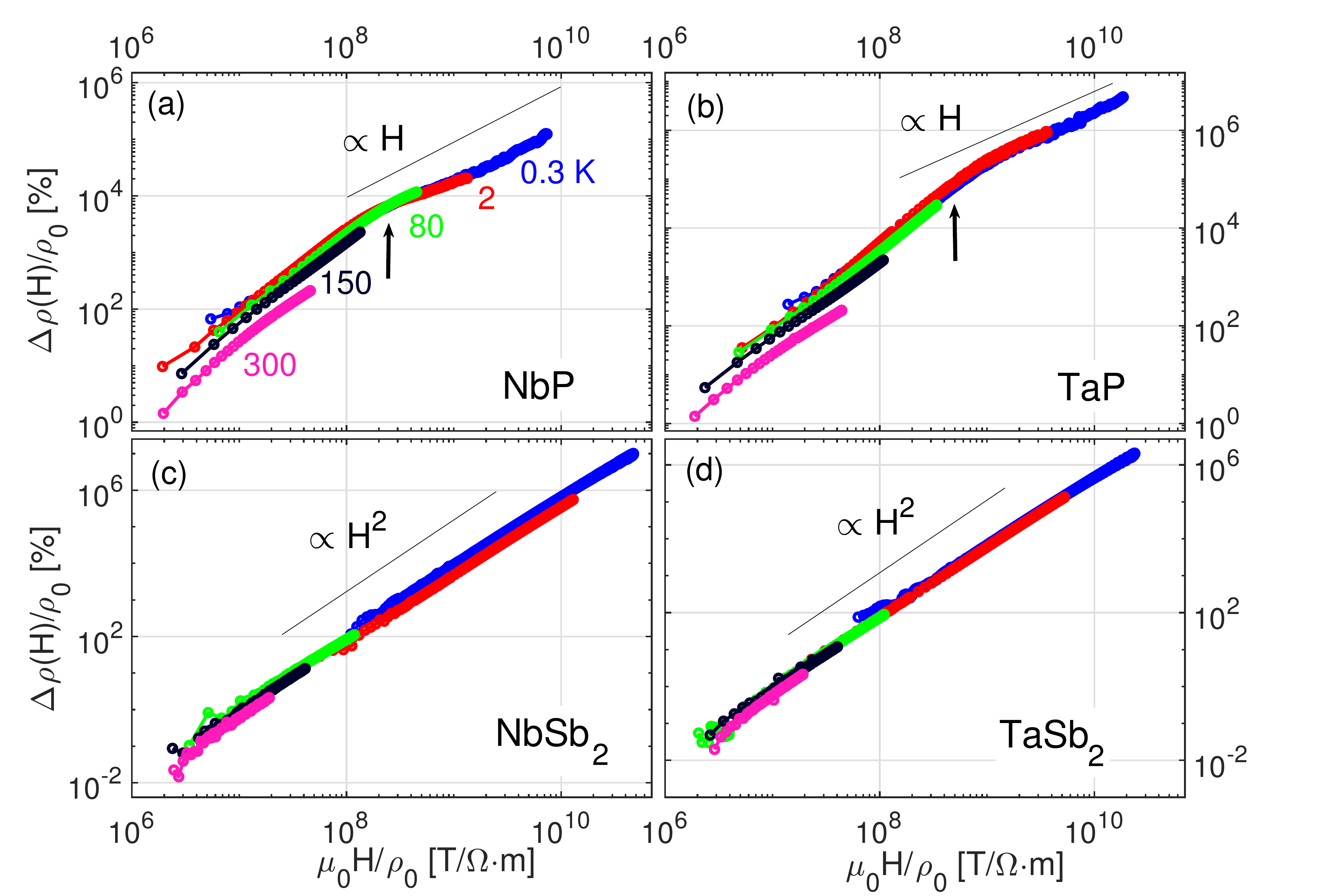}
\caption{\small
Kohler's plots  of phosphides in upper panels (a) NbP, (b) TaP and of antimonates in lower panels (c) \nbsb~ and (d) \tasb. Arrows in upper panels indicate the locations of $H_S$, where $\hallang$ saturates and the MR switches to $H$-linear. }
\label{kohler}
\end{center}
\end{figure}

\black

We now turn to the field dependence of MR. We show Kohler's plots in Fig. \ref{kohler}, and find that  $\Delta\rho/\rho_0 \propto H^\alpha$,%  plotted against $\mu_0H/\rho_0$, 
where $\rho_0 = \rxx(T,H=0)$ --  collapses into a single curve over large $T$ range. In the \black antimonates, we observe $\alpha \approx 2$ in the entire temperature range up to $31$ T. 
In contrast,  in   the \black phosphides the exponent $\alpha$ deviates  from 2  even at low $H$ [$\alpha \approx 1.4\pm0.1$] and switches over  to the  linear-$H$ dependence [$\alpha \approx 1$]  in the  $\mu_0H\ge 8 $ T, where the $\hallang$ approaches a constant value.

\begin{figure}[t!]
\begin{center}
\includegraphics[width=0.9 \linewidth]{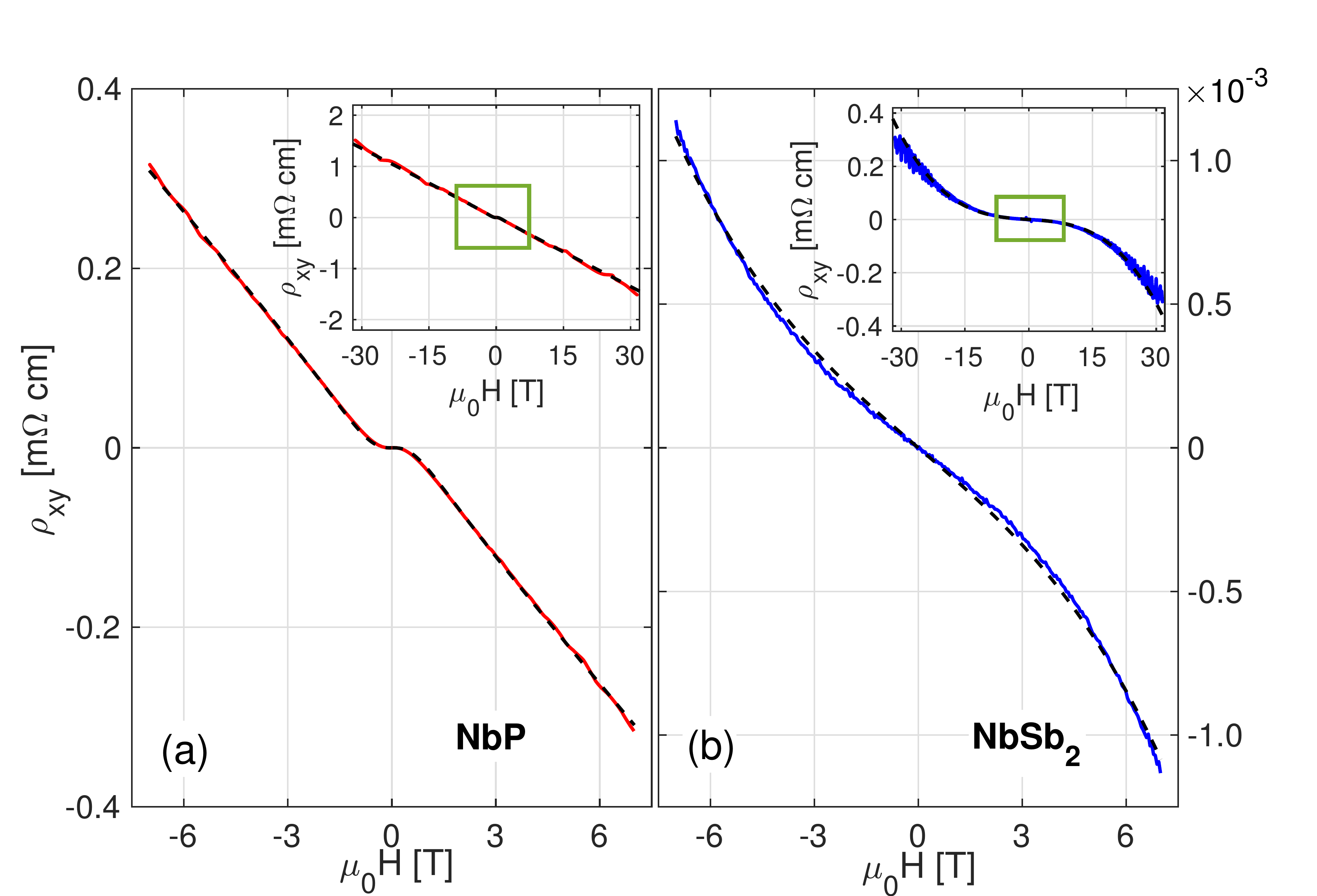}
\caption{\small $\rxy(H)$  as a function of $H$ in (a) NbP and (b)\nbsb, measured at $T=0.3$ K. 
Note the difference of the magnitude of $\rxy$. Broken line in (a) shows a fit to Eq. (\ref{rxyfit}) and (b)  to a two-band model (See SI Appendix  Sec.1 and 2 \cite{SuppInfo}).  Each $\rxy(H)$'s up to  31 T are shown in insets, where the small boxes correspond to the main panels.}
\label{songfit}
\end{center}
\end{figure}

Finally, $\rxy (H)$'s of  NbP and \nbsb~are compared in Fig. \ref{songfit}(a) and (b). The field dependence of \nbsb, shown in the right panel of Fig. \ref{songfit}, is far from linear and a higher power of $H$  becomes more visible with increasing $H$.  $\rxy$ in  TaP and \tasb  showed the similar behavior as  presented in  SI Appendix \cite{SuppInfo} [Fig. S2].

\begin{figure}[t!]
\begin{center}
\includegraphics[width=1 \linewidth]{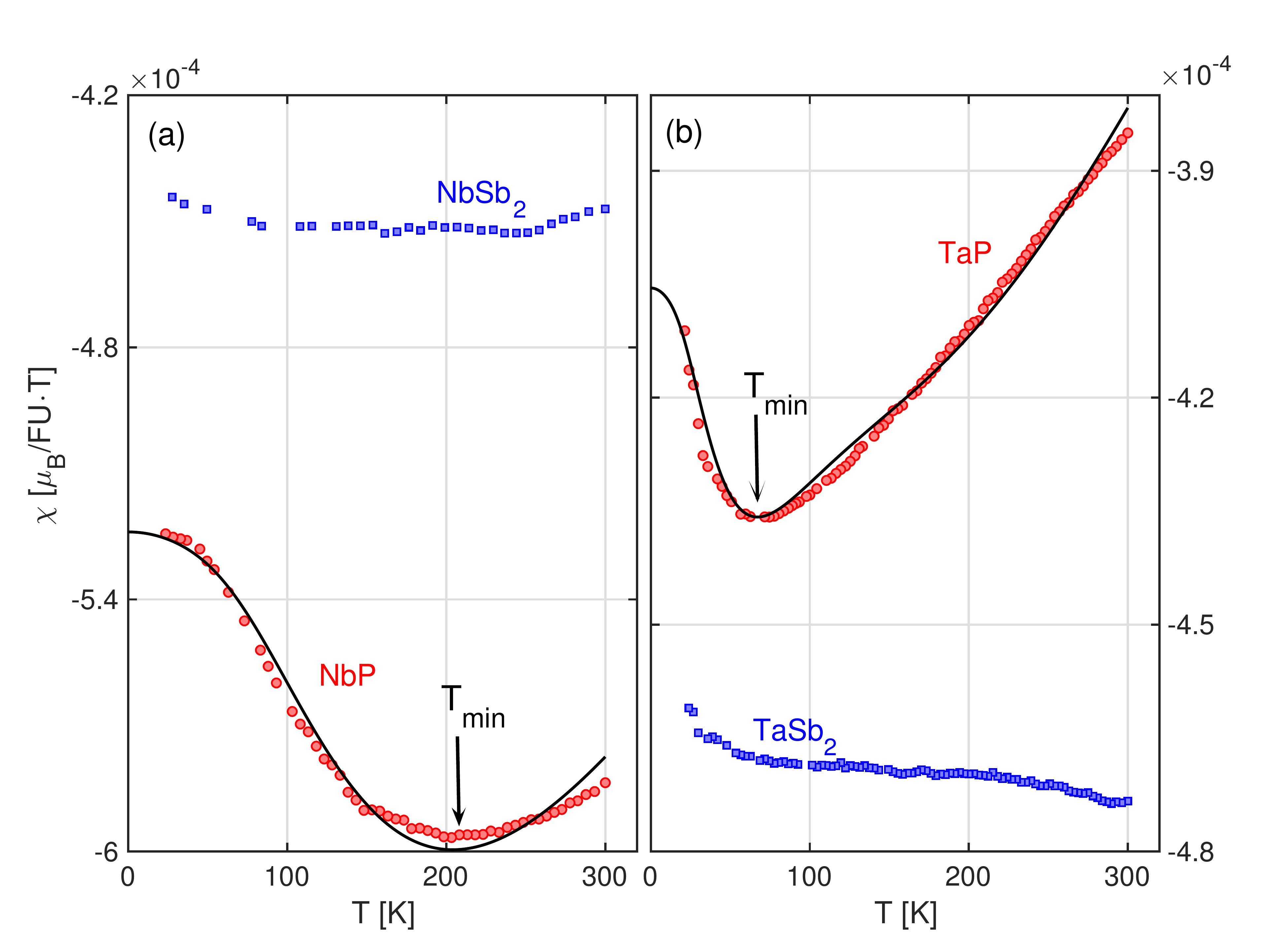}
\caption{\small   $\chi$ vs $T$ in the phosphides (red) and antimonates (blue), measured at $\mu_0H= 1$ T. Solid line is fit to  Eq. (\ref{chifit}), which gives $\mu= -40$ meV for one linear node for NbP and $\mu_1 = 51$ meV and $\mu_2= -11$ meV for two linear nodes for TaP.  Arrows indicate the locations of $T_{\min}$ (See the text). 
 }
\label{chi}
\end{center}
\end{figure}

%In order to track the energy dispersion in the vicinity of the chemical potential, we examine
{\it Magnetic susceptibility}  In Fig.~\ref{chi}, we plot the $T$ dependence of the magnetic susceptibility in the low field limit.
All four samples show negative susceptibilities, corresponding to diamagnetism. 
Fig.~\ref{chi} displays $\chi$ as a function of $T$ for (a) NbP and \nbsb~and (b) TaP and \tasb. 
Both NbP and TaP  have broad yet pronounced minima emerging at  $T_{\min}=203$ K and $T_{\min}=68$ K, respectively. 
%For TaP  the minimum susceptibility [Fig.\ref{chi}] and the resistivity peak under field [Fig. \ref{MRT}(b)] both occur at similar temperatures, which also also close to the temperature where $\tan^2\theta_H$ first becomes appreciable [Fig. \ref{hallang}(c)]. For NbP, these temperatures are within a factor of two, although the agreement is not as close as for TaP
%
%\blue
 For TaP  the minimum susceptibility [Fig.\ref{chi}] and the resistivity peak under field [Fig. \ref{MRT}(b)] both occur at similar temperatures, which are also close to the temperature where $\tan^2\theta_H$ first becomes appreciable [Fig. \ref{hallang}(c)]. For NbP, these temperatures are within a factor of two, although the agreement is not as close as for TaP.
\black 
 
%These minimal temperature   are found to coincide with the onset of increasing $\hallang$ [Fig. 1(c)] for the phosphides. 
On the other hand, $\chi(T)$  for the antimonates remain featureless and mostly constant. 

%%% FIG 4

%%% FIG5

{\emph {Discussion}} 
We begin by discussing the magnetic susceptibility plots shown in Fig.~\ref{chi}. The absence of Pauli paramagnetism in all samples indicates that we do not have spin degenerate bands, and is strong evidence for spin-momentum locking, such that the magnetic susceptibility is dominated by orbital diamagnetism \cite{Blundellbook}. As we now discuss, the additional features can be well explained if we postulate that the phosphides are uncompensated semimetals with a linear dispersion, whereas the antimonates are compensated semimetals. 

We discuss first the phosphides. In particular, we  fit the $\chi(T)$'s of phosphides to  the result of orbital magnetism in a linear dispersion, which is obtained from  energy minimization for the case of a simple linear band crossing (See SI Appendix \cite{SuppInfo})
 \bee
 \chi(T) = C\int_0^{\ep_0} \frac{\eta(\mu_i/k_BT)-\eta(-\mu_i/k_BT)}{\ep}~d\ep,
 \label{chifit}
\ene 
where $\eta$ is the standard Fermi-Dirac distribution function, $\ep_0$ is a cut-off energy,  $\mu$  the  chemical potential measured from $E_F$ to the charge neutrality points and $C$ is a constant in the order of unity.
The data for NbP and TaP is well fit [Fig.\ref{chi}] by  Eq. (\ref{chifit}). For NbP, we find one linear crossing point with $\mu=-40$ meV, which is consistent with one of the reported locations of a Weyl points in NbP  ($-57$ meV and $+5$ meV) \cite{Klotz2016}. 
For TaP, we find two crossing points, $\mu_1= -51$ meV and $\mu_2 = +11$ meV. 
This is again remarkably consistent with the location of a pair of Weyl nodes reported in photoemission data at $-40$ and $+20$ meV \cite{SuYangXu2015}. The magnetic susceptibility data thus strongly suggests that the phosphides should be understood  as uncompensated   semimetals with linear dispersion. \black

This basic conjecture is also consistent with all our observed transport data on the phosphides.
 %\addJS{
 In particular, we note that in the presence of smooth disorder, guiding centers can diffuse in an unusual way to naturally lead to a $H$-linear $\rxx$~\cite{Polyakov1986, Song2015}. This arises when cyclotron radius is smaller than the disorder correlation length (at large enough fields) enabling the guiding center trajectories to become squeezed along the field direction, and exhibit a $\sxx$ that has a dominant $1/H$ dependence~\cite{Song2015}.
% }
 %if we conjecture that the guiding center diffusion scenario \cite{Polyakov1986, Song2015} for magneto-resistance is operative. 
 We note that this scenario (which is only operative for uncompensated semimetals with smooth disorder) also gives an $H$-linear $\rxy$. %Specifically, \addJS{it} %this scenario 
 %predicts \cite{Song2015} that 
%\addJS{
Indeed, writing $\rxy$ in terms of $\sxx$ and $\sxy$~\cite{Song2015}, we obtain%}
\bee
\rxy =\frac{\sxy}{\sxy^2+\sxx^2} = \frac{ne/H}{(ne/H)^2+(b_0/H+\tilde{b}/H^2)^2},
\label {rxyfit}
\ene
where $b_0$ and $\tilde b$ are system specific parameters \cite{Song2015}.  
In the large field limit, this yields $H$-linear $\rxy$,  while in the low field limit a cubic $H$-dependence arises.  
The solid line in Fig. \ref{songfit}(a) is a fit to Eq.(\ref{rxyfit}), and finds good agreement between data and this analytic form, with $ne = 1.9\times 10^6$ c/m$^3$, $b_0=5.6 \times 10^5$ C/m$^3$ and $\tilde b=1.5\times 10^6$ C$\cdot$T/m$^3$. 
%\addJS{
Additionally, these parameters confirm that $1/H$-like dependence dominates $\sxx$ for fields larger than several Tesla, as we find clearly in our our data.
%}.

%This scenario also predicts 
Central to linear MR is a field independent $\hallang$, consistent with observations on the phosphides at fields above $H_S \simeq 8$ T [Fig.\ref{hallang}]. Specifically, %it 
when the cyclotron radius is smaller than the disorder correlation length, Ref.~\cite{Song2015} estimates a $\hallang$ as
\begin{equation}
\hallang \approx \frac{2}{\sqrt{27 \pi}} \left( \frac{\mu}{e V_0}\right)^{3/2}
\end{equation}
where $\mu$ is the chemical potential  and $V_0$ is the disorder strength (typical fluctuation in local chemical potential). Taking the values for $\mu$ from the magnetic susceptibility fit 
(for TaP taking the larger of the $\mu$ values, since the valley with larger Fermi surface will provide most of the carriers), we obtain $V_0 \approx 7$ mV for NbP and $V_0 \approx  10$ mV for TaP. 

%\addJS{
We note that linear MR is expected to disappear for $k_BT \gg V_0$~\cite{Song2015} when inelastic scattering degrades the squeezed trajectories of guiding centers.
%}. 
These are expected to occur above a temperature scale or order of  $\frac{V_0}{k_B} \approx 90$ K and 120 K  for NbP and TaP respectively. In NbP the estimated $T$ scales are consistent with the $T$ scale on which non-monotonicity is observed in Fig. \ref{MRT} (a) and with the temperature dependence of the Kohler plots in Fig. \ref{kohler}(a).  In fact, these two $T$ scales also corresponds to where $\hallang (T)$  measured at $H_S$ begins to rise rapidly, [Fig. 1(c)].  
In TaP,  the observed temperature scale is around 60 K (instead of the expected 120K), however, we remind the reader that it is hard to cleanly separate out $V_0$ scale from MR, because other thermally activated scatterings become important at elevated temperature. 
 %nevertheless, the estimated $T$ scales for $k_BT \gg V_0$ are consistent with the temperature scale on which non-monotonicity is observed in Fig. \ref{MRT}(a-b)and with the temperature dependence of the Kohler plots in Fig. \ref{kohler}(a-b). In fact there two temperature scale correspond to  where $\hallang (T)$  measured at $H_S$ begins to rise rapidly, around 100 K and 60 K for NbP and TaP respectively  [Fig. 1(c)].  
  
\black  We can also extract the disorder correlation length ($\xi$) from the condition that the cyclotron radius ($r_C$) is of the same order as disorder correlation length at $H_S \approx 8$ T, above which $\hallang$ becomes field independent. 
At $H=H_S$,  the $\xi$ is in the order of  $r_C = \frac{mv_F}{eB}$ and  they are estimated to be 26  and 14 nm  for NbP and TaP, respectively,  using for Fermi velocities ($v_F$'s) from  the reported values \cite{Shekhar2015, ChiChengLee2015, Arnold2016}.  It is interesting to note that the values of $\xi$  are consistent with the length scales for defects and stacking faults  that were revealed in TaP \cite{Besara2016}.
\black

%This behavior is closely related to  the  non-monotonic  $\rxx(T)$ for phosphides shown in Fig. 2(a) and (b).  Starting from the low $T$s,  $\rxx' =1/\sxx$ initially decreases rapidly for both phosphides and antimontates, which is attributed to increases in density of sates.  Further increase  of temerpature brought out  thermally-induced scatterings that  drives $\rxx$ larger values.  Prominent non-monontic $\rxx(T)$ is  marked by the maximum $\rxx$  at $T=T_P$, which lies in the vicinity of $V_0$.   The precise origins of this  are presnetly unclear but  we speculate the original of  the peculiar temperature dependence $\rxx$ comes from the disorder potential, which is tied to  the saturated value of $\hallang$. 
  \black

% Fermi velocity ($v_F$)  are  $4.8 \times 10^5$ m/s and $3.5\times 10^5$ m/s \cite{Shekhar2015} and the effective mass ($m*$)'s are $0.08m_e$ and $0.06m_e$

Meanwhile, a field independent $\hallang$ and an $H$-linear $\rxy$ automatically imply an $H$-linear magnetoresistance at high fields, consistent with Fig. \ref{kohler}. Finally, the temperature dependence of $\rxx$ [Fig.\ref{MRT}] can also be understood within this framework. The key point to note is that these systems have linear dispersion with small doping, such that the density of states at the Fermi level is small, $\sim \mu^2$. Increasing the temperature $T$ allows the system to access states within $k_BT$ of the chemical potential, and (since the density of states grows rapidly with energy), greatly enhances the number of states that can participate in transport. We thus conclude that increasing temperature can increase $\sxx$ through this density of states effect, consistent with the observed monotonic decline in $\rxx'$ with increasing temperature  [Fig.\ref{MRT} insets]. 
   %implies that the recurring temperature scale in the electrical transport, $T=T_P$ in $\rho(T)$ and the onset temperature of the increase of $\hallang$, is essentially tied to the chemical potential of the systems with a linear dispersion.

%We thus conclude that 
These, together, establish that all salient observed features of the phosphides can be explained by an
%if we assume that these are 
uncompensated spin orbit locked semimetal with linear dispersion; 
this is corroborated by our linear MR-type magnetotransport expected from guiding center diffusion that is particularly pronounced in semimetals with linear dispersion~\cite{Song2015}. 
Moreover, a systematic combination of thermodynamic and magnetotransport measurements can allow us to extract parameters such as chemical potential (or doping level) and typical disorder strength and correlation length. 
%\addJS{
These enable to make a direct correspondence with a microscopic guiding center description, e.g., identification of fields above which linear MR dominates and  identification of temperature scales below which $\hallang$ becomes large and $H$-independent.
%} 
\black

The antimonates also exhibit diamagnetism, indicating that these materials are also spin-orbit coupled, but their magnetic susceptibility is not well fit by an expression of the type Eq. (\ref{chifit}). Instead, the susceptibility is  mostly temperature-independent, closer to the expectation for  Landau diamagnetism for quadratic bands \cite{Blundellbook}. These materials also exhibit a magnetoresistance that is $\sim H^2$. These facts, as well as the smallness of $\hallang$ in the antimonates, are all well explained  if we postulate that these systems are {\it compensated} semimetals  described by a two band model (See SI appendix Sec. 1 \cite{SuppInfo}). In {\it compensated} semimetals, $\hallang$ is small, as long as  mobilities of carriers remains in the similar range\black , 
and the magnetoresistance is $\sim H^2$, consistent with observation. The magnetoresistance is non-saturating for perfect compensation, but will eventually saturate at a value $\sim 1/\delta n^2$, where $\delta n$ is the difference between electron and hole densities. We ascribe the lack of saturation of MR up to $31$T to the systems being close enough to compensation that we do not hit the saturation value at experimentally accessible fields. 
%\blue
Here we note that, despite high mobilities of both carriers  in antimonates (See SI appendix Table S1 \cite{SuppInfo}), that  satisfy the condition of $\nu B\gg1$,  $\hallang$ is found  much less than unity. This implies  that  the   system effectively remains the limit of $\omega_c \tau \ll 1$  and the MRs of the antimonates should not be  saturated within experimentally accessible field range of this work. 
% and saturation is not expected until $\omega_c \tau \gg 1$, consistent with the picture that MR in the antimonates {\it will} saturate eventually, at higher fields than experimentally accessible. 
\black

The field dependence of $\rxy$ is also informative. With small deviation from perfect compensation, one expects (See SI appendix Sec 3 \cite{SuppInfo}) that $\rxy \sim H$ for systems with linear dispersion, but for quadratic dispersion one expects $\rxy \sim H$ at low fields, with a crossover to $\rxy \sim H^3$ at higher fields. The data in Fig.\ref{songfit} is more consistent with the latter behavior, suggesting that the antimonates should be thought of as compensated semimetals with effectively {\it quadratic} dispersion (i.e. appreciable band curvature on the scale of the doping level). 
This conclusion is also consistent with the magnetic susceptibility data, which is reminiscent of the Landau diamagnetism of quadratic bands. 
%

%The only remaining puzzle is the temperature dependence of the resistivity, which shows a non-monotonic behavior [Fig.\ref{MRT}]. 
%The precise origins of this are presently unclear, but we speculate that the decrease of resistivity with increasing temperature at low temperatures is due to a density of states effect (similar to the phosphides), 
%while the increase of resistivity with increasing temperature at higher temperatures is due to a temperature driven increase in scattering rates. 

{\emph{Summary}} 
We have investigated magneto-transport and magnetic susceptibility of four different semimetals. We find that the combination of susceptibility and magnetotransport measurements allows us to cleanly characterize the non-saturating behavior of MR. The two phosphide materials that we study (NbP and TaP) are well described by a model of uncompensated semimetals with linear dispersion, wherein the magnetoresistance is well described by guiding center diffusion with $\hallang \gg 1$ and field-independent. The combination of measurements that we have made also allows us to extract the disorder strength and disorder correlation length in these materials, as well as the doping level. Meanwhile, the antimonates (\nbsb~and \tasb) are well described as {\it compensated} semimetals governed by a two band model with effectively quadratic bands and $\hallang \ll 1$. 
%In phosphides,  the linear energy dispersion is identified from  the $T$-dependence of the diamagnetic susceptibility.  Simultaneously, we  large and saturated  $\hallang$ in a large field limit serves as  empirical indications that warrant non-saturating linear MR.
% Our results on antimonates, on the other hand,  illustrate the conventional two or multi bands models is  sufficient to describe archetypical magnetotransport properties such as nearly quadratic MR, $T$-independent $\chi(T)$, and a small, non-linear Hall signal. 
The criteria reported here highlight a distinct set of traits for  non-saturating MR and will  serve as a primary touchstone to classify MR phenomena in materials, which in turn, will provide design principles for material platforms  and devices for technological application.

%%%%%%%%%%%%%%%%%%%%%%%%%%%%%%%%%%%%%%%%%%%%%%%%%%%%%%%%%%%%%%
%%%% REFERENCES and Acknowledgement 
%%%%%%%%%%%%%%%%%%%%%%%%%%%%%%%%%%%%%%%%%%%%%%%%%%%%%%%%%%%%%% 

\vspace{0.1in}
\noindent{\bf {Acknowledgment}} This work was supported by the University of Colorado Boulder Office of Research Innovation. We thank David Graf for technical assistance with the high field measurement. High magnetic field data was obtained at the National High Magnetic  Field Laboratory, which is supported by National Science Foundation Cooperative 
Agreement No.~DMR-1157490 and the State of Florida. 
This research was sponsored in part (Y.-P. L. and R.M.N.) by the Army Research Office and was accomplished under Grant Number W911NF-17-1-0482. 
J.C.W.S acknowledges the support of the Singapore National Research Foundation (NRF) under NRF fellowship award NRF-NRFF2016-05.
The views and conclusions contained in this document are those of the authors and should not be interpreted as representing the official policies, either expressed or implied, of the Army Research Office or the U.S. Government. The U.S. Government is authorized to reproduce and distribute reprints for Government purposes notwithstanding any copyright
notation herein.

%\bibliography{lmr}

%\end{document}
%%%%%%%%%%%%%%%%%%%%%%%%%%%%%%%%%%%%%%%%%%%%%%%%%%%%%%%%%%%%%%END %% END %% END %% END %% END %% END %% END %% 
%%%% END %% END %% END %% END %% END %% END %% END %% END %% 
%%% Main Text %%% Main Tex%%% Main Tex%%% Main Tex%%% Main Tex%%% Main Tex%%% Main Tex
%%%%%%%%%%%%%%%%%%%%%%%%%%%%%%%%%%%%%%%%%%%%%%%%%%%%%%%%%%%%%% 

\newpage

%%%%%%%%%%%%%%%%%%%%%%%%%%%%%%%%%%%%%%%%%%%%%%%%%%%%%%%%%%%%%%SuppInfo %% SuppInfo%%SuppInfo %% SuppInfo%%SuppInfo %% SuppInfo
%%%% %%SuppInfo %% SuppInfo%%%%%SuppInfo %% SuppInfo%%%%%%%%%%%%%%%%%%%%%%%%%%%%%%%%%%%%%%%%%%%%%%%%%%%%%%%%%%%%%% 
\newpage
\setcounter{page}{1}
\setcounter{figure}{0}
\renewcommand{\thefigure}{S\arabic{figure}}

\setcounter{equation}{0}
\renewcommand{\theequation}{S\arabic{equation}}

\setcounter{table}{0}
\renewcommand{\thetable}{S\arabic{table}}

\renewcommand*{\citenumfont}[1]{S#1}
\renewcommand*{\bibnumfmt}[1]{[S#1]}

\onecolumngrid

\vskip8mm

\section*{\Large{Supplemental Material}}

\centerline{\bf {Non-saturating large magnetoresistance in semimetals }}
  
\vskip3mm

\centerline{Ian A. Leahy, Yu-ping Lin, Peter E. Siegfried, Andrew C. Treglia, 
Justin C. W. Song,} 

\centerline{Rahul M. Nandkinshore, and Minhyea Lee}

\vskip6mm

\onecolumngrid

\section{S1. Two Band Model Fitting}
%%%%%%% Fig S1
\bfig[!b]
\begin{center}
\includegraphics[width=.8\linewidth]{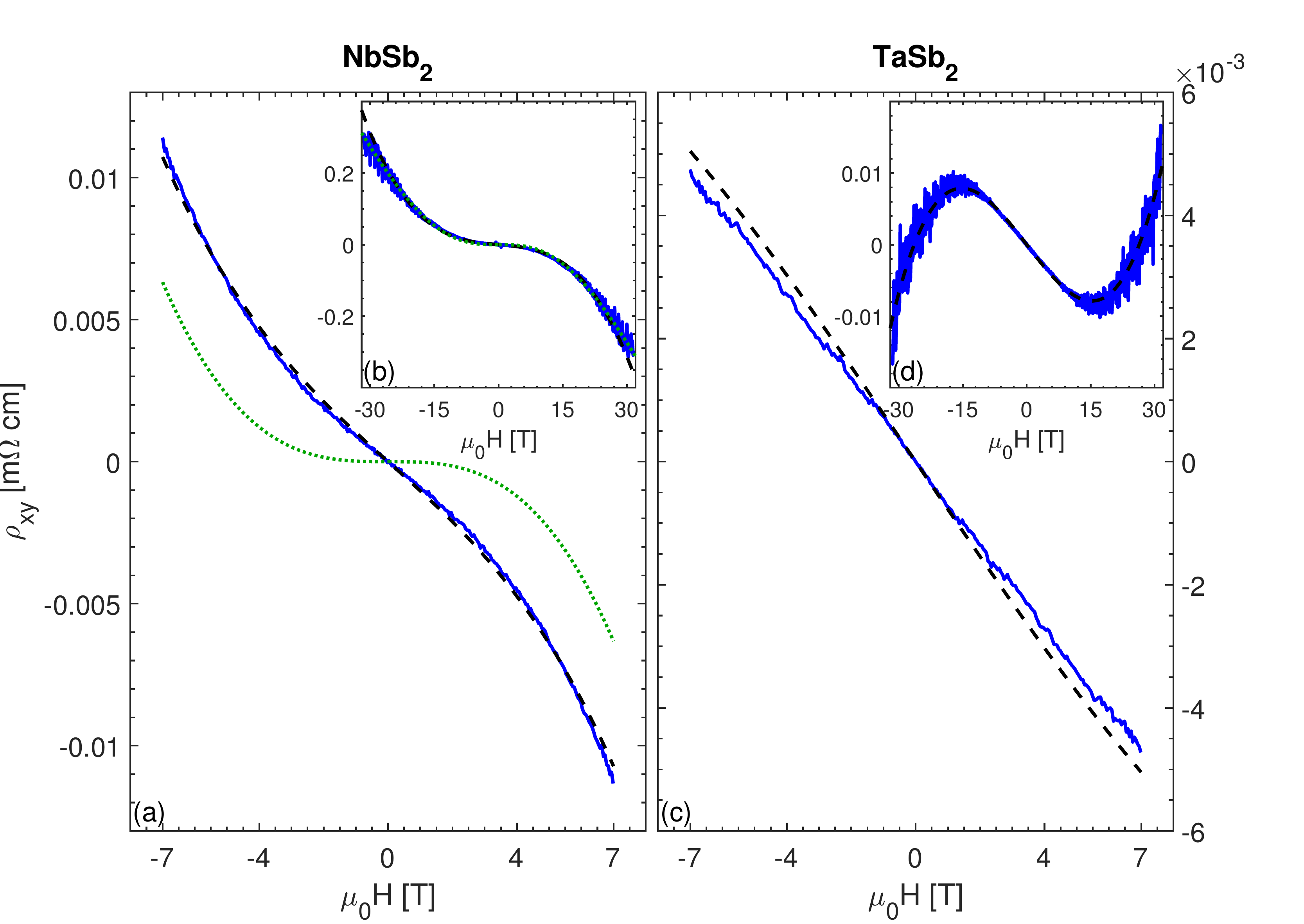}
\caption {\small  $\rho_{xy}$ as a function of $H$ for (a)\nbsb~and (b)\tasb, measured at $T=0.3$ K. The black dashed lines are two band model fits. The green dashed line in (a) is a guiding center diffusion model fit.  The low field dependence is not captured.}
\label{figs1}
\end{center}
\efig
We use the two band model to describe the Hall resistivity of NbSb${}_2$ and TaSb${}_2$:

\begin{eqnarray}
\sigma_{xx}(H)=\sum_i\sigma_i\big[ 1+\left( \frac{\sum_c\sigma_c\mu_c}{\sum_c\sigma_c}  \right)^2H^2  \big]&;&~R_H(H)=\frac{\sum_i\sigma_i\mu_c}{(\sum_c \sigma_c)\cdot\sigma_{xx}(H)} \\
\mathrm{and}~~\sigma_i&=&\frac{N_i e \mu_i}{1+\mu_i^2H^2},
\label{eqnarray}
\end{eqnarray}
where $N_i$ is the carrier density for holes or electrons and $\mu_i$ is the corresponding mobility. The fit results can be seen in Fig. \ref{figs1}, the parameters are given in Table \ref{table1}. From the fit parameters, it is clear that \nbsb~and \tasb~are nearly compensated semimetals: their hole and electron carrier densities differ by fractions of a percent. It is not possible to fit the Hall resistivity for TaP and NbP to a two band model form (see Fig. \ref{figs2}(a)). This is apparent from the low and high field limits of the two band model Hall resistivity. In the low field limit, the two band model exhibits $H$-linear leading order dependence. In the high field limit, it remains $H$-linear. The low field dependence of the phosphides is cubic, not linear, making it impossible to fit with the two band model.

\begin{table}[]
\begin{center}
\begin{tabular}{ccccc} 
\hline\hline
 & N${}_e$ (m${}^{-3}$)  & $\nu_e$ (T$^{-1}$) & N${}_h$ (m${}^{-3}$) & $\nu_h$ (T$^{-1}$)\\ 
\hline\hline
\nbsb& $8.3(1)\times 10^{25}$ & 2.52 & $8.3(0)\times 10^{25}$ & 1.91  \\
\tasb& $2.5(5)\times 10^{26}$   &  4.27 & $2.5(5)\times 10^{26}$ & 2.22 \\
\hline\hline
\end{tabular}
\caption {\small   Two band model fit parameters for \nbsb~and \tasb.  }
\label{tableS1}
\end{center}
\end{table}

\vspace{5mm}
\section{S2. Guiding Center Diffusion Model Fitting}

%%%%%%% Fig S2
\bfig[b!]
\begin{center}
\includegraphics[width=.8\linewidth]{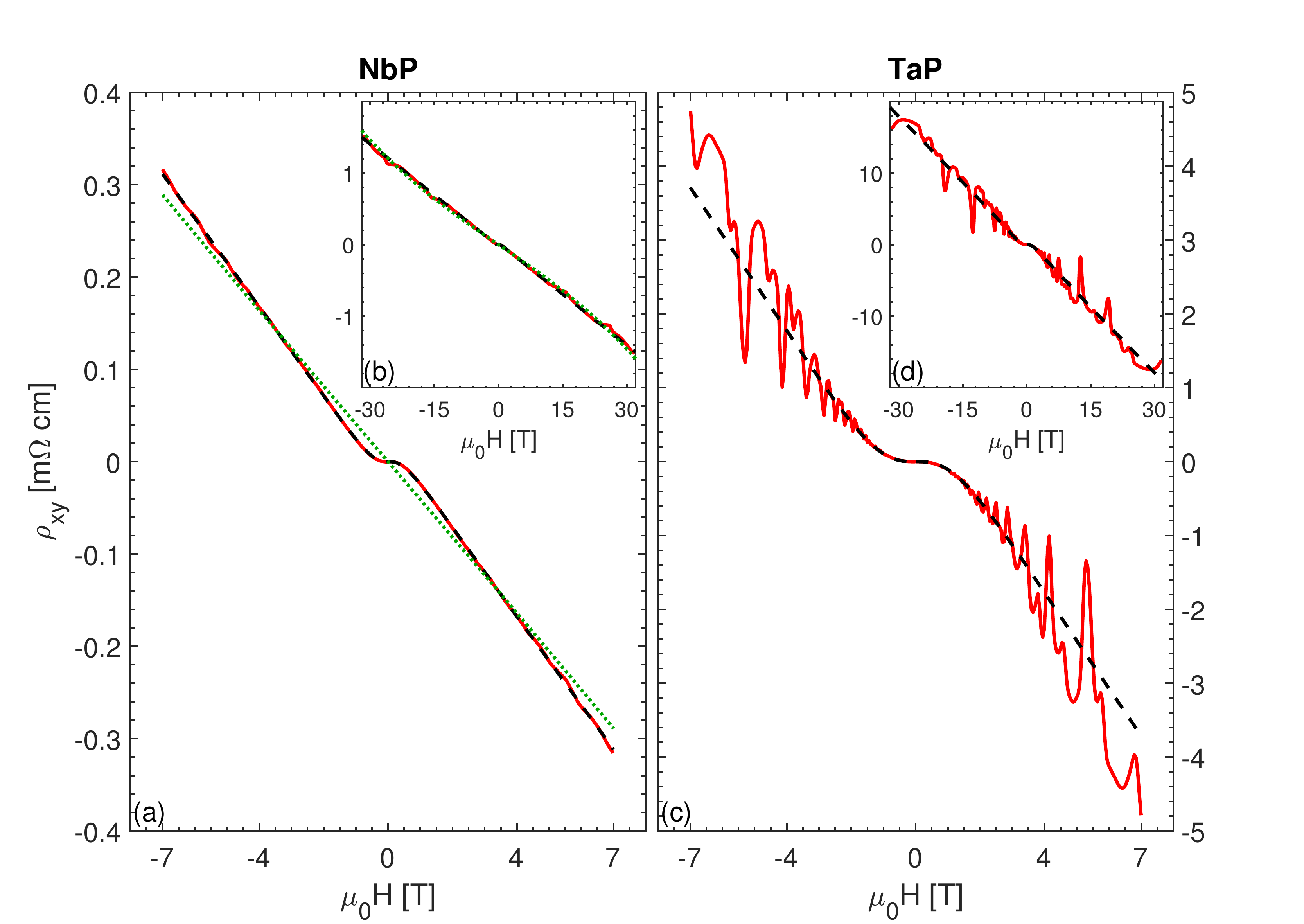}
\caption {\small  $\rho_{xy}$ as a function of $H$ for (a) NbP and (b) TaP, measured at $T=0.3$ K. Black dashed lines are fits to Eq. (\ref{SongEq}). The green dashed line in (a) is a fit to the two band model. The low field dependence is not captured. }
\label{figs2}
\end{center}
\efig

We describe the Hall resistivity of NbP and TaP using the theory of guiding center diffusion in the presence of disorder from Song et. al. \cite{Song}:
\bee
\rxy =\frac{\sxy}{\sxy^2+\sxx^2} = \frac{ne/H}{(ne/H)^2+(b_0/|H|+\tilde{b}/|H|^2)^2},
\label{SongEq}
\ene

where $b_0$ and $\tilde{b}$ are system specific parameters. The fit results to this form are pictured in Fig. \ref{figs2} and the fit parameters are listed in Table \ref{table2}. This form of the Hall resistivity readily reproduces the low field $H^3$ behavior and high field $H$-linear behavior. 

\begin{table}[]
\begin{center}
\begin{tabular}{ccccc} 
\hline\hline
\,\,\, &$\,\,n\cdot e\,$(C/m${}^3$) & $b_0$ (C/m${}^3$) &$\tilde{b}$ (C$\cdot$T/m${}^3$)\\ 
\hline\hline
NbP& $-1.94\times 10^6$ & $5.61\times 10^5$ & $1.47\times 10^6$   \\
TaP& $-1.62\times 10^5$ & $1.72\times 10^4$   &  $3.36\times 10^5$  \\
\hline\hline
\end{tabular}
\caption {\small Fit parameters for the guiding center diffusion model given in Eq. (\ref{SongEq})  }
\label{table2}
\end{center}
\end{table}

\section{S3. Model Hamiltonian}

In this section, we present the electron models with quadratic and linear dispersions. These models serve as the main objects in our later investigations of magnetoresistance and magnetic susceptibility.

As the simplest and the most common model for the descriptions of electronic behavior in materials, the Fermi gas is characterized by the dispersion energy
\beeq
\ve_{\mbf k} = \fr{k^2}{2m}.
\eneq
Here $\mbf k$ is the momentum with magnitude $k=|\mbf k|$, and $m$ is the mass of the electrons. Notice that $\hbar=1$ is assumed. The effective velocity takes the form
\beeq
\label{eq:effvelfg}
\mbf v_{\mbf k} = \delk\ve_{\mbf k} = \fr{\mbf k}{m}.
\eneq
Most of the magnetic properties of Fermi gas have been uncovered. Therefore, the Fermi gas serves as a good benchmark of our calculations before the investigations of Weyl and Luttinger semimetals.

A natural generalization of quadratic dispersion to the two band model framework is provided by the Luttinger semimetal
\beeq
H = \ve_\mrm{NP}+\fr{\vec d(\mbf k)\cdot\vec\G}{2m},
\eneq
where $\ve_\mrm{NP}$ is the energy at the nodal point, and $m$ is the effective mass nearby. The five components of $\vec d(\mbf k)$ are the $l=2$ spherical harmonics
\beeq
\bega
d_1(\mbf k) = \sqrt3k_yk_z,\quad
d_2(\mbf k) = \sqrt3k_zk_x,\quad
d_3(\mbf k) = \sqrt3k_xk_y,\quad
d_4(\mbf k) = \fr{\sqrt3}{2}\lf(k_x^2-k_y^2\ri),\quad
d_5(\mbf k) = \fr{1}{2}\lf(2k_z^2-k_x^2-k_y^2\ri),
\enga
\eneq
and the vector of gamma matrices $\vec\G$ characterizes the $4\times4$ irreducible representations
\beeq
\G^1=\lf(\bear{cc}0&i\s_x\\-i\s_x&0\enar\ri),\quad
\G^2=\lf(\bear{cc}0&i\s_y\\-i\s_y&0\enar\ri),\quad
\G^3=\lf(\bear{cc}0&i\s_z\\-i\s_z&0\enar\ri),\quad
\G^4=\lf(\bear{cc}0&1\\1&0\enar\ri),\quad
\G^5=\lf(\bear{cc}-1&0\\0&1\enar\ri)
\eneq
satisfying the Euclidean Clifford algebra $\{\G^a,\G^b\} = 2\d^{ab}$. The dispersion energies exhibits two quadratic bands
\beeq
\ve_{\mbf k} = \ve_\mrm{NP}\pm\fr{k^2}{2m},
\eneq
with $2$-fold degeneracy on each band. The two energy bands are separated by a nodal point $\ve=\ve_\mrm{NP}$ at $\mbf k=0$ in the Brillouin zone. The effective velocities on the two bands take opposite sign
\beeq
\label{eq:effvellsm}
\mbf v_{\mbf k} = \delk\ve_{\mbf k} = \pm\fr{\mbf k}{m}.
\eneq

The nodal points with linearly dispersing bands play important roles in recent developments of condensed matter physics. A material containing such nodal points with nondegenerate linear bands are called Weyl semimetals. The simplest minimal model for a Weyl point is
\beeq
\label{eq:mmws}
H = \ve_{NP}+v\bsb\s\cdot\mbf k,
\eneq
where $\ve_\mrm{NP}$ is the energy at Weyl point, $v$ is the effective velocity, and $\bsb\s$ is a vector of the Pauli matrices. The model exhibits two linear bands
\beeq
\ve_{\mbf k} = \ve_{NP}\pm vk.
\eneq
The effective velocities of the electrons on the two bands are given by
\beeq
\label{eq:effvelwsm}
\mbf v_{\mbf k} = \delk\ve_{\mbf k} = \pm v\hat k,
\eneq
where $\hat k=\mbf k/k$.

\section{S4. Magnetoresistance}

In this section, we present the calculations of Boltzmann transport theory for the magnetoresistance. The results are exploited in the explanation of quadratic magnetoresistance in nearly compensated systems. The linear magnetoresistance, on the other hand, is described by the guiding center diffusion theory in Ref.~\cite{Song}.

\subsection{S4.1 Boltzmann Transport Theory}

\label{sec:boltztran}

In order to determine the DC magnetoresistance of the models, we calculate the response of electrons to the static electric and magnetic fields in the Boltzmann transport theory \cite{Grosso}. Consider the probability distribution $g_{\mbf k}(\mbf r,t)$, which represents the probability density of an electron carrying momentum $\mbf k$ in position $\mbf r$ and time $t$. Assume that the electric and magnetic fields are infinitesimal, and the disorder in the system is weak. The transport properties of electrons can be described by the semiclassical Boltzmann equation
\beeq
\label{eq:boltzeqori}
\p_tg_{\mbf k}+\dot{\mbf r}\cdot\del g_{\mbf k}+\dot{\mbf k}\cdot\del_{\mbf k}g_{\mbf k} = I_\mrm{coll}[g_{\mbf k}],
\eneq
where $I_\mrm{coll}[g_{\mbf k}]$ is the collision integral. The classical equations of motion are provided as
\beeq
\dot{\mbf r} = \mbf v_{\mbf k},
\quad
\dot{\mbf k} = -e\mbf E-e\dot{\mbf r}\times\mbf B.
\eneq
Imposing the steady state $\p_tg_{\mbf k}=0$ and uniform $\del g_{\mbf k}=0$ conditions, the Boltzmann equation becomes
\beeq
\label{eq:boltzeqdc}
\lf(-e\mbf E-e\mbf v_{\mbf k}\times\mbf B\ri)\cdot\del_{\mbf k}g_{\mbf k}
= I_\mrm{coll}[g_{\mbf k}].
\eneq
We adopt the relaxation time approximation
\beeq
I_\mrm{coll}[g_{\mbf k}] = -\fr{\d g_{\mbf k}}{\tau(\ve_{\mbf k})},
\eneq
where $\tau(\ve)$ is the energy-dependent relaxation time, and $\d g_{\mbf k}=g_{\mbf k}-g_\mrm{eq}$ is the displacement from equilibrium distribution $dg_\mrm{eq}/dt=0$. In the weak-field limit, the equilibrium distribution $g_\mrm{eq}$ is identified with the Fermi-Dirac distribution
\beeq
g_\mrm{eq} = n_F(\ve;\mu) = \fr{1}{e^{\b(\ve-\mu)}+1},
\eneq
where $\b=1/k_BT$ is the inverse temperature and $k_B=1$ is assumed.

With infinitesimal electric field $\mbf E$, the equation Eq.~(\ref{eq:boltzeqdc}) can be linearized with respect to $\mbf E$
\beeq
\lf(-e\mbf E-e\mbf v_{\mbf k}\times\mbf B\ri)\cdot\lf(\delk\ve_{\mbf k}\ri)n_F'(\ve_{\mbf k};\mu)-e\lf(\mbf v_{\mbf k}\times\mbf B\ri)\cdot\delk\d g_{\mbf k}
= -\fr{\d g_{\mbf k}}{\tau(\ve_{\mbf k})}.
\eneq
Since $\delk\ve_{\mbf k}=\mbf v_{\mbf k}$, the equation reduces to a differential equation of $\d g_{\mbf k}$
\beeq
\label{eq:boltzeqdg}
e\tau(\ve_{\mbf k})\mbf v_{\mbf k}\cdot\lf(\mbf B\times\delk\d g_{\mbf k}\ri)-\d g_{\mbf k}
= -e\tau(\ve_{\mbf k})\lf(\mbf E\cdot\mbf v_{\mbf k}\ri)n_F'(\ve_{\mbf k};\mu).
\eneq
We derive an equation for the current density
\beeq
\mbf j
= \int_{\mbf k}g_{\mbf k}\lf(-e\mbf v_{\mbf k}\ri)
= \int_{\mbf k}\d g_{\mbf k}\lf(-e\mbf v_{\mbf k}\ri),
\eneq
where the equilibrium contribution is zero, by multiplying the linearized Boltzmann equation Eq.~(\ref{eq:boltzeqdg}) with $e\mbf v_{\mbf k}$ and integrate over the momentum space
\beeq
\label{eq:boltzeqdcj}
e^2\intvk\tau(\ve_{\mbf k})\mbf v_{\mbf k}\mbf v_{\mbf k}\cdot\lf(\mbf B\times\delk\d g_{\mbf k}\ri)+\mbf j = -e^2\intvk\tau(\ve_{\mbf k})n_F'(\ve_{\mbf k};\mu)\mbf v_{\mbf k}\mbf v_{\mbf k}\cdot\mbf E.
\eneq
We are interested in the systems with the energy bands isotropic about $\mbf k=0$. The effective velocity can therefore be defined as $\mbf v_{\mbf k} = v_k\hat k$. For any vector $\mbf V=V\hat k_n$, an angular integral in momentum space can be simplified
\beeq
\int d\O\hat k\hat k\cdot\mbf V = 2\pi V\hat k_n\hcint d\t\sin\t\cos^2\t = \fr{4\pi}{3}\mbf V,
\eneq
which implies
\beeq
\int d\O\hat k\hat k = \fr{4\pi}{3}.
\eneq
The right hand side of Eq.~(\ref{eq:boltzeqdcj}) is therefore simplified
\begin{align}
-e^2\intvk\tau(\ve_{\mbf k})n_F'(\ve_{\mbf k};\mu)\mbf v_{\mbf k}\mbf v_{\mbf k}\cdot\mbf E
&= \lf[-\fr{1}{(2\pi)^3}\fr{4\pi}{3}e^2\pint dkk^2\tau(\ve_{\mbf k})n_F'(\ve_{\mbf k};\mu)v_k^2\ri]\mbf E
\\
&= \lf[-\fr{1}{(2\pi)^3}\fr{4\pi}{3}e^2\pint d\ve k^2(\ve)\tau(\ve)v_k(\ve)n_F'(\ve;\mu)\ri]\mbf E
,
\end{align}
and the proportionality to the electric field $\mbf E$ can be observed. The integral on the left hand side of Eq.~(\ref{eq:boltzeqdcj}) becomes
\begin{align}
e^2\intvk\tau(\ve_{\mbf k})\mbf v_{\mbf k}\mbf v_{\mbf k}\cdot\lf(\mbf B\times\delk\d g_{\mbf k}\ri)
&= e^2\intvk\tau(\ve_{\mbf k})\lf(\fr{v_k}{k}\mbf k\ri)\lf(\fr{v_k}{k}\mbf k\ri)\cdot\lf(\mbf B\times\delk\d g_{\mbf k}\ri)
\\
&= \hat ae^2\intvk\tau(\ve_{\mbf k})\lf(\fr{v_k}{k}\ri)^2k_a\ve_{bcd}k_bB_c\p_{k_d}\d g_{\mbf k}
\\
&= -\hat a\ve_{bcd}B_ce^2\intvk\p_{k_d}\lf[\tau(\ve_{\mbf k})\lf(\fr{v_k}{k}\ri)^2k_ak_b\ri]\d g_{\mbf k}
\\
&= -\hat a\ve_{bcd}B_ce^2\intvk\lf\{
\fr{k_d}{k}\p_k\lf[\tau(\ve_{\mbf k})\lf(\fr{v_k}{k}\ri)^2\ri]k_ak_b
+\tau(\ve_{\mbf k})\lf(\fr{v_k}{k}\ri)^2\lf(\d_{da}k_b+\d_{db}k_a\ri)
\ri\}\d g_{\mbf k}
\\
&= -\hat a\ve_{bca}B_ce^2\intvk\tau(\ve_{\mbf k})\lf(\fr{v_k}{k}\ri)^2k_b\d g_{\mbf k} \\
&= \mbf B\times e^2\intvk\tau(\ve_{\mbf k})\fr{v_k}{k}\mbf v_{\mbf k}\d g_{\mbf k}.
\end{align}
With the above calculations, the equation for current density Eq.~(\ref{eq:boltzeqdcj}) reduces to
\beeq
\label{eq:dcjeq}
\mbf B\times e^2\intvk\tau(\ve_{\mbf k})\fr{v_k}{k}\mbf v_{\mbf k}\d g_{\mbf k}+\mbf j
= \lf[-\fr{1}{(2\pi)^3}\fr{4\pi}{3}e^2\pint d\ve k^2(\ve)\tau(\ve)v_k(\ve)n_F'(\ve;\mu)\ri]\mbf E.
\eneq
The remaining task is to deal with the integrals containing the distributions $n_F(\ve;\mu)$ and $\d g_{\mbf k}$.

For simplicity, we deal with the zero temperature limit $T=0$. The deviation of probability distribution $\d g_{\mbf k}$ from equilibrium is localized near the chemical potential $\ve=\mu$, and $n_F'(\ve;\mu)=-\d(\ve-\mu)$ reduces to a delta function. Assume that the magnetic field lies in the $z$ direction $\mbf B=B\hat z$. The equation for current density Eq.~(\ref{eq:dcjeq}) becomes
\beeq
\lf[-eB\tau(\mu)\fr{v_{k_\mu}}{k_\mu}(\hat z\times)+1\ri]\mbf j
= \lf[\fr{1}{(2\pi)^3}\fr{4\pi}{3}k^2(\mu)e^2\tau(\mu)v_k(\mu)\ri]\mbf E,
\eneq
where the outer product operator with respect to $z$ direction is defined as
\beeq
(\hat z\times)=\lf(\bear{ccc}0&-1&0\\1&0&0\\0&0&0\enar\ri).
\eneq
The right hand side can be further simplified
\beeq
\lf[\fr{1}{(2\pi)^3}\fr{4\pi}{3}k^3(\mu)e^2\tau(\mu)\fr{v_k(\mu)}{k(\mu)}\ri]\mbf E
= \lf[n_ee^2\tau(\mu)\fr{v_k(\mu)}{k(\mu)}\ri]\mbf E,
\eneq
where
\beeq
n_e = \fr{1}{(2\pi)^3}\fr{4\pi}{3}k^3(\mu)
\eneq
is the electron density. We arrive at the DC current equation 
\beeq
\lf[-eB\tau(\mu)\fr{v_{k_\mu}}{k_\mu}(\hat z\times)+1\ri]\mbf j
= \lf[n_ee^2\tau(\mu)\fr{v_k(\mu)}{k(\mu)}\ri]\mbf E.
\eneq
With the relation $\mbf E=\rho\mbf j$, the resistivity tensor can be identified as
\beeq
\label{eq:boltzresten}
\rho = \lf[n_ee^2\tau(\mu)\fr{v_k(\mu)}{k(\mu)}\ri]^{-1}
\lf(\bear{ccc}1&eB\tau(\mu)\fr{v_{k_\mu}}{k_\mu}&0\\-eB\tau(\mu)\fr{v_{k_\mu}}{k_\mu}&1&0\\0&0&1\enar\ri)
.
\eneq
The exact form of the resistivity tensor $\rho$ depends on the dispersion law and the relaxation time in each model.

\subsection{S4.2 Fermi Gas}

For the Fermi gas, the effective velocity Eq.~(\ref{eq:effvelfg}) implies $v_k/k=1/m$. The electron density $n_e$ is doubled since each momentum mode can contain two electrons. With the assumption of consant relaxation time $\tau(\ve)=\tau$, the resistivity tensor takes the form
\beeq
\label{eq:reselec}
\rho_e = \lf(\fr{n_ee^2\tau}{m}\ri)^{-1}
\lf(\bear{ccc}1&\o_c\tau&0\\-\o_c\tau&1&0\\0&0&1\enar\ri),
\eneq
where $\o_c=eB/m$ is the cyclotron frequency. The result is consistent with that of the Drude model \cite{Ashcroft}, where the longitudinal and Hall resistivities are
\beeq
\rho_{xx} = \fr{m}{n_ee^2\tau},\quad
\rho_{xy} = \fr{B}{n_ee}.
\eneq

\subsection{S4.3 Luttinger Semimetal}

\subsubsection{S4.3a Single Nodal Point}

In the zero temperature limit, the magnetoresistance only acquires contributions from the Fermi surface. When the chemical potential is higher than the nodal point energy $\mu>\ve_\mrm{NP}$, the resistivity tensor of the electron pocket is the same as that of the Fermi gas Eq.~(\ref{eq:reselec}). On the other hand, if the chemical potential is lower than the nodal point energy $\mu<\ve_\mrm{NP}$, the hole carriers dominates the magnetoresistance. Denote the hole density by $n_h$. With $e_h = e$ and $m_h = m$, the resistivity tensor of hole pocket takes the form
\beeq
\label{eq:reshole}
\rho_h = \lf(\fr{n_ee^2\tau}{m}\ri)^{-1}
\lf(\bear{ccc}1&-\o_c\tau&0\\\o_c\tau&1&0\\0&0&1\enar\ri).
\eneq

\subsubsection{S4.3b Two Nodal Points}

When the system is composed of two Luttinger points at different energies $\ve_\mrm{NP1}$ and $\ve_\mrm{NP2}$, more interesting characteristics of magnetoresistance can show up. Different locations of chemical potential can lead to different kinds of magnetoresistance. The simplest case happens when the chemical potential is above or below both nodal points. Since the charge carriers in the two pockets are of the same kind, the result is qualitatively similar to the Fermi gas. Amazing things happen when the chemical potential lies between the two nodal points $\ve_\mrm{NP1}<\mu<\ve_\mrm{NP2}$ \cite{Ashcroft}. An electron pocket and a hole pocket show up in the system, which contribute to the electron and hole densities $n_e$ and $n_h$. The effective electron density is defined as $n_\mrm{eff}=n_e-n_h$. Assume that the effective mass near the two Luttinger points are the same. We calculate the magnetoresistance of the model in this setup. 

The conductivity tensors for both pockets are derived by inversing the resistivity tensors Eq.~(\ref{eq:reselec}) and Eq.~(\ref{eq:reshole})
\beeq
\s_e
= \fr{n_ee^2\tau}{m}
\lf(\bear{ccc}\fr{1}{1+\o_c^2\tau^2}&-\fr{\o_c\tau}{1+\o_c^2\tau^2}&0\\\fr{\o_c\tau}{1+\o_c^2\tau^2}&\fr{1}{1+\o_c^2\tau^2}&0\\0&0&1\enar\ri),
\quad
\s_h=
\fr{n_he^2\tau}{m}
\lf(\bear{ccc}\fr{1}{1+\o_c^2\tau^2}&\fr{\o_c\tau}{1+\o_c^2\tau^2}&0\\-\fr{\o_c\tau}{1+\o_c^2\tau^2}&\fr{1}{1+\o_c^2\tau^2}&0\\0&0&1\enar\ri).
\eneq
To determine the total conductivity tensor of the system, we add the two conductivity tensors $\s_e$ and $\s_h$
\beeq
\s=\fr{e^2\tau}{m}
\lf(\bear{ccc}(n_e+n_h)\fr{1}{1+\o_c^2\tau^2}&-n_\mrm{eff}\fr{\o_c\tau}{1+\o_c^2\tau^2}&0\\
n_\mrm{eff}\fr{\o_c\tau}{1+\o_c^2\tau^2}&(n_e+n_h)\fr{1}{1+\o_c^2\tau^2}&0\\
0&0&n_e+n_h\enar\ri),
\eneq
and the total resistivity tensor is then derived
\beeq
\rho=\fr{m}{e^2\tau}
\lf(\bear{ccc}C(n_e+n_h)&Cn_\mrm{eff}\o_c\tau&0\\
-Cn_\mrm{eff}\o_c\tau&C(n_e+n_h)&0\\
0&0&\fr{1}{n_e+n_h}\enar\ri).
\eneq
The coefficient $C$ is given by
\beeq
C
= \lf(1+\o_c^2\tau^2\ri)\lf[\lf(n_e+n_h\ri)^2+\lf(n_\mrm{eff}\o_c\tau\ri)^2\ri]^{-1}
.
\eneq
With the resistivity tensor, the longitudinal and Hall resistivities can be identified as
\beeq
\rho_{xx}
= \fr{m}{e^2\tau}\lf(1+\o_c^2\tau^2\ri)\fr{n_e+n_h}{\lf(n_e+n_h\ri)^2+\lf(n_\mrm{eff}\o_c\tau\ri)^2},
\quad
\rho_{xy}
= \fr{m}{e^2\tau}\lf(1+\o_c^2\tau^2\ri)\o_c\tau\fr{n_\mrm{eff}}{\lf(n_e+n_h\ri)^2+\lf(n_\mrm{eff}\o_c\tau\ri)^2}.
\eneq

Assume that the system is not compensated $n_\mrm{eff}\neq0$. The resistivities are dominated by different scaling forms when the system experiences different strengths of magnetic field. When the magnetic field is small, the resistivities acquires the approximate forms
\beeq
\label{eq:lutttwolowb}
\rho_{xx} \apx \fr{m}{e^2\tau}\fr{1}{n_e+n_h}\lf\{1+\lf[1-\lf(\fr{n_\mrm{eff}}{n_e+n_h}\ri)^2\ri]\o_c^2\tau^2\ri\},\quad
\rho_{xy} \apx \fr{B}{e}\fr{n_\mrm{eff}}{(n_e+n_h)^2}.
\eneq
The longitudinal resistivity has a quadratic scaling in the magnetic field $B$, and the Hall resistivity is linear in $B$. As the magnetic field increases, the scaling form of resistivity changes. The quartic and cubic scaling dominate the longitudinal and Hall resistivities in the moderate magnetic field regime, respectively. In the high magnetic field regime, the resistivies becomes
\beeq
\rho_{xx} \apx \fr{m}{e^2\tau}\fr{n_e+n_h}{n_\mrm{eff}^2},\quad
\rho_{xy} \apx \fr{B}{n_\mrm{eff}e}.
\eneq
The longitudinal resistivity saturates in the limit of high magnetic field, and the Hall resistivity becomes linear in $B$.

In the nearly compensated case $n_\mrm{eff}\ll n_e,n_h$, the low magnetic field scalings Eq.~(\ref{eq:lutttwolowb}) survive larger magnetic field due to the suppression of high order terms. More impressive phenomena happen when the system is perfectly compensated $n_e=n_h=n$, $n_\mrm{eff}=0$. In this case, the longitudinal and Hall resistivities are
\beeq
\rho_{xx} = \fr{m}{2ne^2\tau}\lf(1+\o_c^2\tau^2\ri),\quad
\rho_{xy}=0.
\eneq
While the longitudinal resistivity increases unboundedly with increasing magnetic field in a quadratic form, the Hall resistivity vanishes exactly.

\subsection{S4.4 Weyl Semimetal}

\subsubsection{S4.4a Single Nodal Point}

In the case of Weyl semimetal, the magnetoresistance is different from the Fermi gas due to the difference in dispersion law. Assume that the chemical potential $\mu$ is above the Weyl point energy $\mu>\ve_\mrm{NP}$. For a short-range impurity scattering potential, the energy-dependent relaxation time $\tau(\ve)$ is determined from the first Born approximation \cite{Burkov2011PRB}
\beeq
\fr{1}{\tau(\ve)} = 2\pi\g\nu(\ve),
\eneq
where $\g$ is a parameter depending on the impurities, and $\nu(\ve)$ is the density of states
\beeq
\nu(\ve) = \fr{1}{(2\pi)^3}4\pi k^2\fr{dk}{d\ve} = \fr{(\ve-\ve_\mrm{NP})^2}{2\pi^2v^3}.
\eneq
With
\beeq
\tau(\mu)\fr{v_k(\mu)}{k(\mu)}
= \fr{\pi v^3}{\g(\mu-\ve_\mrm{NP})^2}\fr{v}{k(\mu)}
= \fr{\pi v^2}{\g k^3(\mu)}
= \fr{\pi v^2[1/(2\pi)^3](4\pi/3)}{\g[1/(2\pi)^3](4\pi/3)k^3(\mu)}
= \fr{v^2}{6\pi\g n_e}
= \fr{\k}{n_e},
\eneq
where $\k=v^2/6\pi\g$ is defined, the resistivity tensor at zero temperature is calculated from Eq.~(\ref{eq:boltzresten})
\beeq
\label{eq:weylresten}
\rho_e = \lf(\k e^2\ri)^{-1}
\lf(\bear{ccc}1&\fr{eB\k}{n_e}&0\\-\fr{eB\k}{n_e}&1&0\\0&0&1\enar\ri).
\eneq
The longitudinal and Hall resistivities can be read from the tensor $\rho$
\beeq
\rho_{xx} = \fr{1}{\k e^2},\quad
\rho_{xy} = \fr{B}{n_ee}.
\eneq

When the chemical potential is lower than the Weyl point energy $\mu<\ve_\mrm{NP}$, the hole pocket dominates the magnetoresistance. For the hole carriers, the charge $e_h=e$ is opposite to the electron charge $-e$. Hence, the resistivity tensor takes the form
\beeq
\label{eq:weylholeresten}
\rho_h = \lf(\k e^2\ri)^{-1}
\lf(\bear{ccc}1&-\fr{eB\k}{n_h}&0\\\fr{eB\k}{n_h}&1&0\\0&0&1\enar\ri).
\eneq

\subsubsection{S4.4b Two Nodal Points}

As in the case of Luttinger semimetal, we consider a system which consists of two Weyl points at different energies $\ve_\mrm{NP1}$ and $\ve_\mrm{NP2}$. When the chemical potential is above or below both Weyl points $\mu>\ve_\mrm{NP1},\ve_\mrm{NP2}$, the charge carriers in the two pockets are of the same kind, and the results are similar to those of the single Weyl point. As the chemical potential lies between the two nodal points $\ve_\mrm{NP1}<\mu<\ve_\mrm{NP2}$, there exist an electron pocket and a hole pocket in the system, which provides nontrivial features of magnetoresistance. The electron and hole densities are denoted by $n_e$ and $n_h$, and the effective electron density is defined as $n_\mrm{eff}=n_e-n_h$. We assume the same impurity parameter $\g$ for the two Weyl points.

The conductivity tensors for the two pockets are obtained by inversing the corresponding resistivity tensors Eq.~(\ref{eq:weylresten}) and Eq.~(\ref{eq:weylholeresten})
\beeq
\s_e
= \lf(\k e^2\ri)
\lf(\bear{ccc}\fr{1}{1+(eB\k/n_e)^2}&-\fr{eB\k/n_e}{1+(eB\k/n_e)^2}&0\\\fr{eB\k/n_e}{1+(eB\k/n_e)^2}&\fr{1}{1+(eB\k/n_e)^2}&0\\0&0&1\enar\ri),
\quad
\s_h=\lf(\k e^2\ri)
\lf(\bear{ccc}\fr{1}{1+(eB\k/n_h)^2}&\fr{eB\k/n_h}{1+(eB\k/n_h)^2}&0\\-\fr{eB\k/n_h}{1+(eB\k/n_h)^2}&\fr{1}{1+(eB\k/n_h)^2}&0\\0&0&1\enar\ri).
\eneq
We derive the total conductivity tensor by adding these two tensors
\begin{align}
\s
&= \k e^2
\lf(\bear{ccc}\fr{1}{1+(eB\k/n_e)^2}+\fr{1}{1+(eB\k/n_h)^2}&-\fr{eB\k/n_e}{1+(eB\k/n_e)^2}+\fr{eB\k/n_h}{1+(eB\k/n_h)^2}&0\\
\fr{eB\k/n_e}{1+(eB\k/n_e)^2}-\fr{eB\k/n_h}{1+(eB\k/n_h)^2}&\fr{1}{1+(eB\k/n_e)^2}+\fr{1}{1+(eB\k/n_h)^2}&0\\
0&0&2\enar\ri)
\\
&= \k e^2
\lf(\bear{ccc}\fr{2+(eB\k)^2(1/n_e^2+1/n_h^2)}{[1+(eB\k/n_e)^2][1+(eB\k/n_h)^2]}&-\fr{(eB\k)(1/n_e-1/n_h)[1-(eB\k)^2/n_en_h]}{[1+(eB\k/n_e)^2][1+(eB\k/n_h)^2]}&0\\
\fr{(eB\k)(1/n_e-1/n_h)[1-(eB\k)^2/n_en_h]}{[1+(eB\k/n_e)^2][1+(eB\k/n_h)^2]}&\fr{2+(eB\k)^2(1/n_e^2+1/n_h^2)}{[1+(eB\k/n_e)^2][1+(eB\k/n_h)^2]}&0\\
0&0&2\enar\ri)
\\
&= \k e^2
\lf(\bear{ccc}\fr{2+(eB\k)^2(1/n_e^2+1/n_h^2)}{[1+(eB\k/n_e)^2][1+(eB\k/n_h)^2]}&\fr{(eB\k)(n_\mrm{eff}/n_en_h)[1-(eB\k)^2/n_en_h]}{[1+(eB\k/n_e)^2][1+(eB\k/n_h)^2]}&0\\
-\fr{(eB\k)(n_\mrm{eff}/n_en_h)[1-(eB\k)^2/n_en_h]}{[1+(eB\k/n_e)^2][1+(eB\k/n_h)^2]}&\fr{2+(eB\k)^2(1/n_e^2+1/n_h^2)}{[1+(eB\k/n_e)^2][1+(eB\k/n_h)^2]}&0\\
0&0&2\enar\ri)
.
\end{align}
The resistivity tensor is the inverse of conductivity tensor $\s$
\beeq
\rho \apx \fr{1}{\k e^2}
\lf(\bear{ccc}C\lf\{2+(eB\k)^2\lf(\fr{1}{n_e^2}+\fr{1}{n_h^2}\ri)\ri\}&-CeB\k\fr{n_\mrm{eff}}{n_en_h}\lf[1-\fr{(eB\k)^2}{n_en_h}\ri]&0\\
CeB\k\fr{n_\mrm{eff}}{n_en_h}\lf[1-\fr{(eB\k)^2}{n_en_h}\ri]&C\lf\{2+(eB\k)^2\lf(\fr{1}{n_e^2}+\fr{1}{n_h^2}\ri)\ri\}&0\\
0&0&\fr{1}{2}\enar\ri),
\eneq
where the coefficient $C$ is
\beeq
C
= \lf[1+\lf(\fr{eB\k}{n_e}\ri)^2\ri]\lf[1+\lf(\fr{eB\k}{n_h}\ri)^2\ri]
\lf\{\lf[2+(eB\k)^2\lf(\fr{1}{n_e^2}+\fr{1}{n_h^2}\ri)\ri]^2+(eB\k)^2\lf(\fr{n_\mrm{eff}}{n_en_h}\ri)^2\lf[1-\fr{(eB\k)^2}{n_en_h}\ri]^2\ri\}^{-1}
.
\eneq
The longitudinal and Hall resistivities are given by
\begin{align}
\rho_{xx}
&= \fr{1}{\k e^2}\lf[1+\lf(\fr{eB\k}{n_e}\ri)^2\ri]\lf[1+\lf(\fr{eB\k}{n_h}\ri)^2\ri]
\fr{2+(eB\k)^2(1/n_e^2+1/n_h^2)}{[2+(eB\k)^2(1/n_e^2+1/n_h^2)]^2+(eB\k)^2(n_\mrm{eff}/n_en_h)^2[1-(eB\k)^2/n_en_h]^2}
,\\
\rho_{xy}
&= -\fr{1}{\k e^2}\lf[1+\lf(\fr{eB\k}{n_e}\ri)^2\ri]\lf[1+\lf(\fr{eB\k}{n_h}\ri)^2\ri]
\fr{eB\k(n_\mrm{eff}/n_en_h)[1-(eB\k)^2/n_en_h]}{[2+(eB\k)^2(1/n_e^2+1/n_h^2)]^2+(eB\k)^2(n_\mrm{eff}/n_en_h)^2[1-(eB\k)^2/n_en_h]^2}.
\end{align}

Assume that the system is not compensated $n_\mrm{eff}\neq0$. The resistivities are dominated by different scaling forms when the system lies in different magnetic field regimes. When the magnetic field is small, the longitudinal resistivity acquires the approximate form
\begin{align}
\rho_{xx}
&\apx \fr{1}{2\k e^2}\lf[1+\lf(\fr{eB\k}{n_e}\ri)^2\ri]\lf[1+\lf(\fr{eB\k}{n_h}\ri)^2\ri]\lf[1+\fr{1}{2}(eB\k)^2\lf(\fr{1}{n_e^2}+\fr{1}{n_h^2}\ri)\ri]
\lf\{1-(eB\k)^2\lf[\fr{1}{n_e^2}+\fr{1}{n_h^2}+\fr{1}{4}\lf(\fr{n_\mrm{eff}}{n_en_h}\ri)^2\ri]\ri\}
\\
&\apx \fr{1}{2\k e^2}
\lf\{1+\fr{1}{2}(eB\k)^2\lf[\fr{1}{n_e^2}+\fr{1}{n_h^2}-\fr{1}{2}\lf(\fr{n_\mrm{eff}}{n_en_h}\ri)^2\ri]\ri\}
,
\end{align}
which has a quadratic scaling in the magnetic field $B$. The Hall resistivvity is linear in $B$
\beeq
\rho_{xy}
\apx -\fr{B}{4e}\fr{n_\mrm{eff}}{n_en_h}.
\eneq
In the moderate magnetic field regime, the higher order scalings dominate the resistivities. In particular, the cubic scaling shows up and reverses the sign of the Hall resistivity when the magnetic field is large enough. In the high magnetic field regime, the longitudinal resistivity saturates
\beeq
\rho_{xx} \apx \fr{1}{\k e^2}\fr{1/n_e^2+1/n_h^2}{(n_\mrm{eff}/n_en_h)^2},
\eneq
and the Hall resistivity is linear in the magnetic field $B$ with a reversed sign
\beeq
\rho_{xy} \apx \fr{B}{n_\mrm{eff}e}.
\eneq

In the nearly compensated situation $|n_\mrm{eff}|\ll n_e,n_h$, the small field scaling survives larger magnetic field due to the suppression of high order terms. More impressive phenomena happen when the system is perfectly compensated $n_e=n_h=n$, $n_\mrm{eff}=0$. In this case, the longitudinal and Hall resistivities are
\beeq
\rho_{xx} = \fr{1}{2\k e^2}\lf[1+\lf(\fr{eB\k}{n}\ri)^2\ri],\quad
\rho_{xy}=0.
\eneq
The longitudinal resistivity increases quadratically and is unbounded, while the Hall resistivity vanishes.

\section{S5. Magnetic Susceptibility}

In this section, we calculate the magnetic susceptibilities of the models that have been encountered \cite{Koshino}. When a magnetic field $\mbf B=B\hat z$ is introduced to these models, the energy spectra exhibit discrete energy levels, known as the Landau levels. These quantized energies are denoted by $\ve_{nk_z}$, where $n$ is the Landau level index and $k_z$ is the $z$ component of momentum. Notice that we only deal with the orbital contribution to the susceptibility, which usually provides diamagnetic feature. The paramagnetic behavior resulting from Zeeman splitting of spin degeneracy is neglected in our investigations. Such treatment is valid especially for the Weyl semimetal, since there is no spin degeneracy to be split.

With the free energy
\beeq
f
= -\fr{1}{\b}\fr{1}{2\pi l_B^2}\int_{k_z}
\sum_n\sum_s\ln\lf[1+e^{-\b(\ve_{nk_z}-\mu)}\ri]
,
\eneq
where $l_B=1/\sqrt{eB}$ is the magnetic length and $s$ is the spin index, the susceptibility can be identified as the second order derivative
\begin{align}
\chi
&= -\fr{\p^2f}{\p B^2} \\
&= \fr{e}{2\pi\b}\int_{k_z}\sum_n\sum_s
\fr{\p^2}{\p B^2}\lf\{B\ln\lf[1+e^{-\b(\ve_{nk_z}-\mu)}\ri]\ri\}
\\
&= \fr{e}{2\pi\b}\int_{k_z}\sum_n\sum_s
\lf\{
2\fr{\p}{\p B}\ln\lf[1+e^{-\b(\ve_{nk_z}-\mu)}\ri]
+B\fr{\p^2}{\p B^2}\ln\lf[1+e^{-\b(\ve_{nk_z}-\mu)}\ri]
\ri\}
\\
\label{eq:susgenform}
&= -\fr{e}{2\pi}\int_{k_z}\sum_n\sum_s
\lf[
2\fr{\p\ve_{nk_z}}{\p B}n_F(\ve_{nk_z};\mu)
+B\fr{\p^2\ve_{nk_z}}{\p B^2}n_F(\ve_{nk_z};\mu)
+B\lf(\fr{\p\ve_{nk_z}}{\p B}\ri)^2n_F'(\ve_{nk_z};\mu)
\ri]
.
\end{align}
For simplicity, we restrict the studies to the zero field limit $B=0$ at finite temperature $T>0$. The factor $B$ provides potential elimination of the last two terms in the square bracket. However, there may exist divergence when $k_z=0$ and $B=0$, which can lead to nonvanishing contributions. We will show that this issue does not happen for the models we consider.

\begin{figure}[b]
\centering
\includegraphics[scale = 0.85]{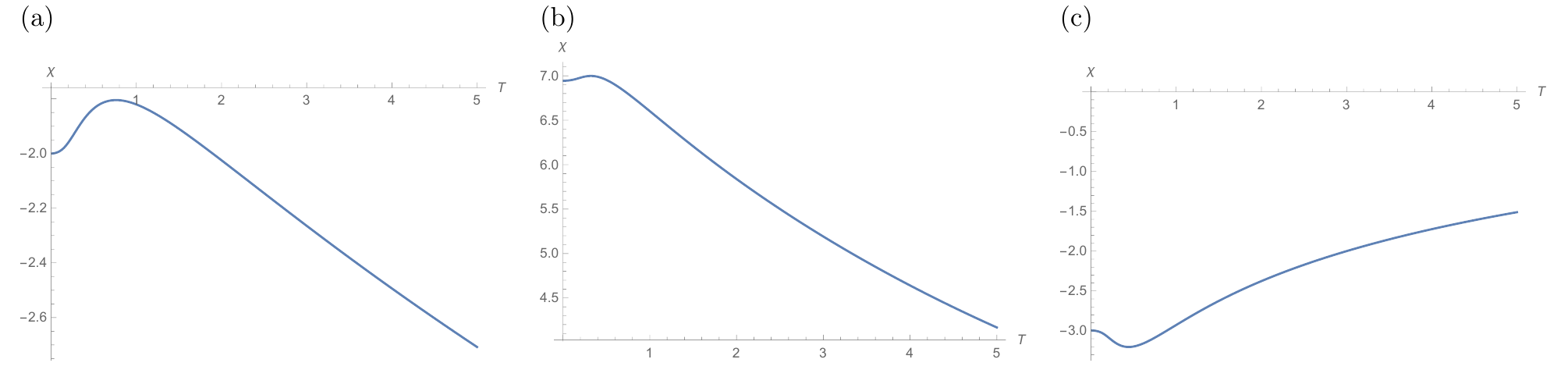}
\caption{\label{fig:sus} Zero field susceptibility of (a) Fermi gas (b) Luttinger semimetal (c) Weyl semimetal. The energy cutoff is set as $\ve_0=20$ for the Luttinger and Weyl semimetals. The other parameters are set as $1$.}
\end{figure}

\subsection{S5.1 Fermi Gas}

We first calculate the susceptibility of Fermi gas. The Landau level energy is
\beeq
\ve_{nk_z} = \o_c\lf(n+\fr{1}{2}\ri)+\fr{k_z^2}{2m},\quad
n=0,1,2,\dots.
\eneq
Notice that each orbital mode carries two electron states. Since the last two terms in the square bracket in Eq.~(\ref{eq:susgenform}) always vanish as $B\rar0$, the zero field susceptibility only acquires a contribution from the first term
\beeq
\chi
= -\fr{e^2}{\pi m}\int_{k_z}\lf[\sum_{n=0}^\infty\lf(2n+1\ri)\ri]
n_F(k_z^2/2m;\mu).
\eneq
The value of the infinite series can be determined from the Ramanujon summation $1+2+3+\dots=-1/12$. With the identity $\sum_{n=0}^\infty(2n+1)=\sum_{n=0}^\infty n-\sum_{n=0}^\infty 2n=1/12$, the zero field susceptibility becomes
\beeq
\chi
= -\fr{e^2}{12\pi m}\int_{k_z}n_F(k_z^2/2m;\mu),
\eneq
and can be written in terms of energy integral
\beeq
\chi = -\fr{e^2}{12\pi^2}\pint d\ve\fr{1}{(2m\ve)^{1/2}}n_F(\ve;\mu).
\eneq
This result is identical to the one in Ref.~\onlinecite{Pathria} except for the factor $2$ of the degeneracy in each orbital mode.

The susceptibility exhibits a turning over behavior as the temperature increases [Fig.~\ref{fig:sus}(a)]. To inspect this phenomenon, a change of variable $x=\ve/T$ is applied to the integral
\beeq
\chi
= -\fr{e^2\mu^{1/2}}{12\pi^2[2m(\mu/T)]^{1/2}}\pint dx\fr{1}{x^{1/2}}\fr{1}{e^{x-\mu/T}+1}
.
\eneq
The zero field susceptibility can be regarded as a function of $\mu/T$ when the chemical potential $\mu$ is fixed. When the temperature is low $\mu/T\gg1$, an increasing quadratic scaling can be determined from the Sommerfeld expansion. In the high temperature limit $\mu/T\ll1$, a square root scaling $\chi\sim-T^{1/2}$ scaling appears. The turning point is located at a fixed $\mu/T_T$. Therefore, the turning point temperature $T_T$ is proportional to the chemical potential $\mu$. This feature provides a potential way of determining the Fermi level in the materials.

\subsection{S 5.2 Luttinger Semimetal}

For the Luttinger semimetal, the Landau level extends to the negative energy regime
\beeq
\ve_{\pm nk_z} = \ve_\mrm{NP}\pm\lf[\o_c\lf(n+\fr{1}{2}\ri)+\fr{k_z^2}{2m}\ri].
\eneq
The derivative $\p\ve/\p B$ in Eq.~(\ref{eq:susgenform}) acquires a negative sign in the negative energy Landau levels. Therefore, the zero field susceptibility for Luttinger semimetal takes the form
\beeq
\chi
= -\fr{e^2}{12\pi^2}\pint d\ve\fr{1}{(2m\ve)^{1/2}}\lf[n_F(\ve;\mu-\ve_\mrm{NP})-n_F(-\ve;\mu-\ve_\mrm{NP})\ri]
.
\eneq
The integral for the negative energy part is divergent. To extract the relevant characteristics near the Fermi level and the nodal point energy, a cutoff $\ve_0\gg|\mu-\ve_\mrm{NP}|$ of the energy scale is introduced to the integral
\beeq
\chi = -\fr{e^2}{12\pi^2}\int_0^{\ve_0} d\ve\fr{1}{(2m\ve)^{1/2}}\lf[n_F(\ve;\mu-\ve_\mrm{NP})-n_F(-\ve;\mu-\ve_\mrm{NP})\ri]
.
\eneq
The square bracket can be reduced to a more explicit form
\beeq
\label{eq:fermidiff}
n_F(\ve;\mu-\ve_\mrm{NP})-n_F(-\ve;\mu-\ve_\mrm{NP})
=-\fr{\sinh\b\ve}{\cosh\b(\mu-\ve_\mrm{NP})+\cosh\b\ve}.
\eneq
Notice that the negative energy part renders the integral positive, indicating a paramagnetic feature of susceptibility in the Luttinger semimetal. However, this effect might be an artifact of the chosen energy cutoff $\ve_0$. Whether the paramagnetic feature is realistic requires further investigations.

Similar to the Fermi gas, the susceptibility manifests quadratic increase in the low temperature regime $|\mu-\ve_\mrm{NP}|/T\gg1$ [Fig.~\ref{fig:sus}(b)]. When the temperature is high $|\mu-\ve_\mrm{NP}|/T\ll1$, the approximation $\cosh\b(\mu-\ve_\mrm{NP})\apx1$ implies
\beeq
\chi
\apx \fr{e^2\mu^{1/2}}{12\pi^2[2m(\mu/T)]^{1/2}}\int_0^{\b\ve_0} dx\fr{1}{x^{1/2}}\fr{\sinh x}{1+\cosh x}
.
\eneq
The integral is dominated by the upper bound $\b\ve_0$ of the integral, indicating a scaling $\chi\sim T^{-1}$ in the high temperature regime. A turning point shows up at certain temperature $T_T$. Since the upper bound $\b\ve_0$ is effectively infinite at moderate temperature, the zero field susceptibility can be regarded as a function of $(\mu-\ve_\mrm{NP})/T$ as in the case of Fermi gas. Therefore, the turning point temperature $T_T$ is proportional to the deviation of chemical potential from the nodal point energy $\mu-\ve_\mrm{NP}$.

\subsection{S5.3 Weyl Semimetal}

We first calculate the Landau level energy spectrum of the Weyl semimetal \cite{Koshino}. With the definition of gauge field $\mbf B=\curl\mbf A$, the Hamiltonian Eq.~(\ref{eq:mmws}) becomes
\beeq
\label{eq:mmwsmf}
H = \ve_\mrm{NP}+v\bsb\s\cdot\bsb\pi.
\eneq
The kinetic momentum $\bsb\pi=\mbf k+e\mbf A$ satisfies the commutation relation $[\pi_x,\pi_y]=-ieB$. To diagonalize the Hamiltonian, we consider the annihilation and creation operators
\beeq
a = \fr{1}{\sqrt{2eB}}\lf(\pi_x-i\pi_y\ri),\quad
a^\dag = \fr{1}{\sqrt{2eB}}\lf(\pi_x+i\pi_y\ri)
\eneq
with the commutation relation $[a,a^\dag]=1$. The Hamiltonian Eq.~(\ref{eq:mmwsmf}) can be expressed in terms of $a$ and $a^\dag$
\beeq
H = \ve_\mrm{NP}+v\sqrt{2eB}\lf(\s^+a+\s^-a^\dag\ri)+v\s^zk_z.
\eneq

For $|n|\geq1$, the eigenstates at $\pm n$-th level can be expressed with the basis $\{\phi_{n-1}\ket{\uar},\phi_n\ket{\dar}\}$, where $\phi_n$'s are the Landau level wavefunctions of 2D free electrons
\beeq
a\phi_n = \sqrt n\phi_{n-1},\quad a^\dag\phi_n = \sqrt{n+1}\phi_{n+1}.
\eneq
The matrix representation of Hamiltonian in this basis
\beeq
H = \ve_\mrm{NP}+\lf(\bear{cc}vk_z&v\sqrt{2neB}\\v\sqrt{2neB}&-vk_z\enar\ri)
\eneq
indicates that the energy spectrum is
\beeq
\ve_{\pm nk_z} = \ve_\mrm{NP}\pm v\sqrt{2neB+k_z^2},\quad n\geq1.
\eneq
When $n=0$, the zero-th Landau level is $\phi_0\ket{\dar}$. The energy is given by
\beeq
\ve_{0k_z} = \ve_\mrm{NP}-vk_z.
\eneq

Since the energy of zero-th Landau level $\ve_{0k_z}$ is independent of magnetic field $B$, only the Landau levels with $|n|\geq1$ contributes. Each term in the square bracket in Eq.~(\ref{eq:susgenform}) has to be checked. The first term does not vanish when $B\rar0$
\begin{align}
&\lf.
-\fr{e}{\pi}\sum_{a=\pm}\int_{k_z}\sum_{n=1}^\infty
\lf(av\fr{ne}{\sqrt{2neB+k_z^2}}\ri)n_F(\ve_{ank_z};\mu)
\ri|_{B=0}
\\
&= -\fr{e^2v^2}{\pi}\int_{k_z}\sum_{a=\pm}\lf(\sum_{n=1}^\infty n\ri)
\fr{1}{av|k_z|}n_F(av|k_z|;\mu-\ve_\mrm{NP})
\\
&= \fr{e^2v^2}{12\pi}\int_{k_z}\sum_{a=\pm}\fr{1}{av|k_z|}n_F(av|k_z|;\mu-\ve_\mrm{NP})
\\
&= \fr{e^2v}{12\pi^2}\int_0^\infty d\ve\fr{1}{\ve}\lf[n_F(\ve;\mu-\ve_\mrm{NP})-n_F(-\ve;\mu-\ve_\mrm{NP})\ri]
.\label{eqsus}
\end{align}
To check the second integral, we consider the calculation
\begin{align}
\iint dk_zB\fr{\p^2\ve_{ank_z}}{\p B^2}
&= \iint dk_zB\lf(-av\fr{n^2e^2}{\lf(2neB+k_z^2\ri)^{3/2}}\ri) \\
&= -\fr{avn^2e^2B}{(2neB)^{3/2}}\iint dk_z\fr{1}{[1+(k_z/\sqrt{2neB})^2]^{3/2}} \\
&= -\fr{avn^2e^2B}{2neB}\int_{-\pi/2}^{\pi/2} d\t\sec^2\t\fr{1}{\sec^3\t},\quad  \tan\t=k_z/\sqrt{2neB} \\
&= -\fr{avne}{2}\int_{-\pi/2}^{\pi/2} d\t\cos\t \\
&= -avne.
\end{align}
In the $B=0$ limit, the integrand vanishes as $k_z\neq0$ and diverges at $k_z=0$. The structure implies a delta function form in the limit $B\rar0$
\beeq
\lim_{B\rar0}B\fr{\p^2\ve_{ank_z}}{\p B^2} = -avne\d(k_z).
\eneq
Hence, the second integral vanishes
\begin{align}
\lf.-\fr{e}{2\pi}\sum_{a=\pm}\int_{k_z}\sum_{n=1}^\infty
B\fr{\p^2\ve_{ank_z}}{\p B^2}n_F(\ve_{ank_z};\mu)\ri|_{B=0}
&= -\fr{e}{2\pi}\sum_{a=\pm}\int_{k_z}\sum_{n=1}^\infty
\lf[-avne\d(k_z)\ri]n_F(\ve_\mrm{NP}+av|k_z|;\mu)
\\
&= \fr{e^2v}{4\pi^2}\sum_{a=\pm}a\lf(\sum_{n=1}^\infty n\ri)
n_F(\ve_\mrm{NP};\mu)
\\
&= 0.
\end{align}
The third term
\beeq
B\lf(\fr{\p\ve_{ank_z}}{\p B}\ri)^2
= B\lf(v^2\fr{n^2e^2}{2neB+k_z^2}\ri)
\eneq
vanishes for $k_z\neq0$ and acquires a finite value $v^2ne/2$ at $k_z=0$ in the $B=0$ limit. This functional form implies that the integral vanishes when $B=0$. We conclude that the zero field susceptibility of Weyl semimetal takes the form \cite{Mikitik}
\beeq
\chi
= \fr{e^2v}{12\pi^2}\int_0^{\ve_0} d\ve\fr{1}{\ve}\lf[n_F(\ve;\mu-\ve_\mrm{NP})-n_F(-\ve;\mu-\ve_\mrm{NP})\ri]
.\label{eqsus}
\eneq
The energy cutoff $\ve_0\gg|\mu-\ve_\mrm{NP}|$ excludes the contribution provided by states away from the Weyl point.

We discuss the dependence of zero field susceptibility on temperature  [Fig.~\ref{fig:sus}(c)]. In the low temperature regime $|\mu-\ve_\mrm{NP}|/T\gg1$, a decreasing quadratic scaling can be verified with Sommerfled expansion. When the temperature is high $|\mu-\ve_\mrm{NP}|/T\ll1$, the approximation $\cosh\b(\mu-\ve_\mrm{NP})\apx1$ in Eq.~(\ref{eq:fermidiff}) implies
\beeq
\chi
\apx -\fr{e^2v}{12\pi^2}\int_0^{\b\ve_0} dx\fr{1}{x}\fr{\sinh x}{1+\cosh x}
.
\eneq
The integral is dominated by the upper bound $\b\ve_0$ of integral, indicating a logarithmic scaling $\chi\sim \ln T$ in the high temperature regime. There is a turning point temperature $T_T$ where the zero field susceptibility turns over. Similar to the case of Luttinger semimetal, the zero field susceptibility can be regarded as a function of $(\mu-\ve_\mrm{NP})/T$ when the chemical potential $\mu$ is fixed. Since the zero field susceptibility turns over at a fixed $(\mu-\ve_\mrm{NP})/T_T$, we can conclude that the turning point $T_T$ is proportional to the difference between chemical potential and Weyl point energy $\mu-\ve_\mrm{NP}$.

\end{document}